 \documentclass[pmlr,twocolumn,10pt]{jmlr} % W&CP article

\usepackage{booktabs}
\usepackage{siunitx}
\usepackage{enumitem}

 \usepackage[switch]{lineno}

\theorembodyfont{\upshape}
\theoremheaderfont{\scshape}
\theorempostheader{:}
\theoremsep{\newline}

\jmlrvolume{297}
\jmlryear{2025}
\jmlrworkshop{Machine Learning for Health (ML4H) 2025} % W&CP title

 \title[TempoQL]{TempoQL: A Readable, Precise, and Portable Query System for Electronic Health Record Data}

 \author{%
  \Name{Ziyong Ma} \Email{ziyongm@andrew.cmu.edu}\\
    \addr Carnegie Mellon University, USA
    \AND
  \Name{Richard D. Boyce} \Email{rdb20@pitt.edu}\\
    \addr University of Pittsburgh, USA
    \AND
  \Name{Adam Perer} \Email{adamperer@cmu.edu}\\
  \Name{Venkatesh Sivaraman} \Email{venkats@cmu.edu}\\
  \addr Carnegie Mellon University, USA
 }

\begin{document}

\maketitle

\begin{abstract}
Electronic health record (EHR) data is an essential data source for machine learning for health, but researchers and clinicians face steep barriers in extracting and validating EHR data for modeling. 
Existing tools incur trade-offs between expressivity and usability and are typically specialized to a single data standard, making it difficult to write temporal queries that are ready for modern model-building pipelines and adaptable to new datasets. 
This paper introduces \textbf{TempoQL}, a Python-based toolkit designed to lower these barriers. 
TempoQL provides a simple, human-readable language for temporal queries; support for multiple EHR data standards, including OMOP, MEDS, and others; and an interactive notebook-based query interface with optional large language model (LLM) authoring assistance. 
% Together, these features reduce technical overhead, improve collaboration between clinicians and data scientists, and mitigate the risk of silent errors. 
Through a performance evaluation and two use cases on different datasets, we demonstrate that TempoQL simplifies the creation of cohorts for machine learning while maintaining precision, speed, and reproducibility. % love it
\end{abstract}

\paragraph*{Data and Code Availability}
This paper uses the MIMIC-IV~\citep{johnson_mimic-iv_2023}, eICU~\citep{pollard2018eicu}, and EHRSHOT~\citep{wornow2023ehrshot} datasets as examples throughout.
The code for TempoQL is available at \url{https://github.com/cmudig/tempo-ql} and can be installed via PyPI (\texttt{pip install tempo-ql}).

\paragraph*{Institutional Review Board (IRB)}
This research did not require IRB approval.
% This initial paragraph is \textbf{mandatory}. If your research requires IRB
% approval or has been designated by your IRB as Not Human Subject
% Research, then for the camera-ready version of the paper, you must
% provide IRB information (and at the time of submission for review, you
% can say that this IRB information will be provided if the paper is
% accepted). If your research does not require IRB approval, then you
% must state this to be the case. 

\section{Introduction}
\label{sec:intro}

\begin{figure*}
    \centering
    \includegraphics[width=\linewidth]{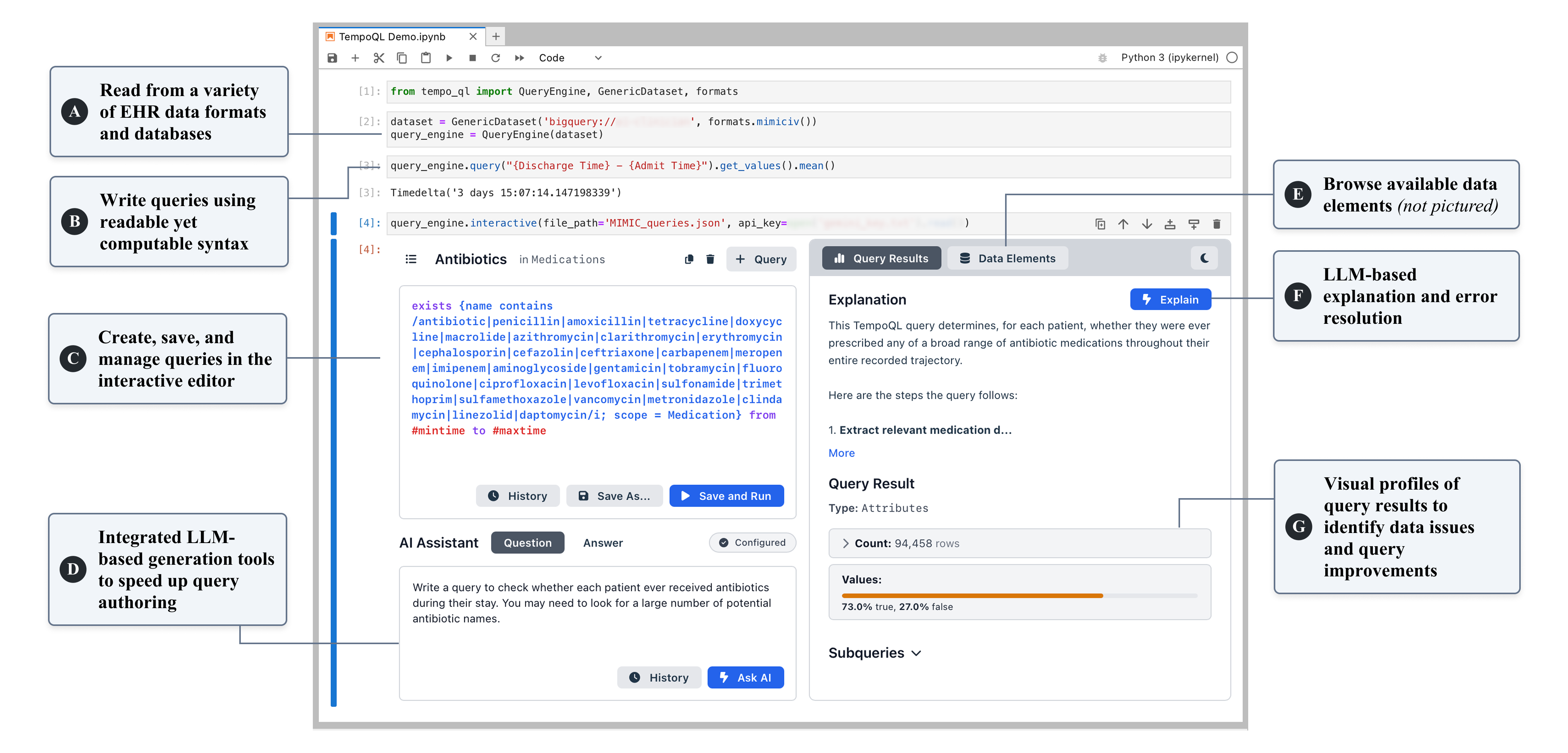}
    \caption{Overview of TempoQL's query authoring process.}
    \label{fig:ui-overview}
\end{figure*}

% Healthcare stands at the forefront of domains where machine learning can have transformative impact. 
From predicting disease progression and personalizing treatment plans to optimizing hospital resource allocation, machine learning models built on electronic health record (EHR) data can help save lives, reduce costs and increase access to care. 
However, despite this immense potential, exploring, transforming, and validating EHR data remains a fundamental barrier to developing high-quality models.
Typical EHR databases can contain thousands of clinical concepts and millions of observations~\citep{johnson_machine_2016}, making it challenging to robustly identify constructs of interest.
Additionally, EHR data is often encoded in heterogeneous formats across institutions and health care data collaboratives (e.g., OHDSI, PCORnet, N3C, and others), so porting modeling pipelines across different datasets is often cumbersome~\citep{arnrich_medical_2024}.
Even when datasets are stored in consistent formats, the same data extraction processes may not work because hospital systems may code the same observations in different ways ~\citep{johnson_machine_2016}, which necessitates careful validation at each stage of model building.
% hold the key to building robust models, but they are often fragmented across institutions, encoded in heterogeneous standards like OMOP or FHIR. 
Due to these challenges, researchers and data scientists in ML for healthcare (ML4H) currently face steep barriers in extracting the features and predictive variables that are needed to build robust ML models. 

Existing tools for querying EHR data offer different balances of flexibility and accessibility. 
On one side, database query languages and frameworks such as the HL7 Clinical Quality Language\footnote{\url{https://cql.hl7.org/index.html}} provide fine-grained control and expressive power but require substantial technical expertise to author.
For example, seemingly simple temporal queries needed for ML model building, such as ``extract the patient's last blood pressure measurement every hour,'' can require complex code to handle the possibility of multiple related concepts and sparse or missing values.
This technical complexity can prevent domain experts from providing their insight into the data and modeling procedures~\citep{sivaraman_tempo_2025,schwartz_clinician_2021}.
Graphical user interfaces such as cohort builders or drag-and-drop EHR tools~\citep{krause_supporting_2016, zhang2015iterative} abstract away some of this complexity, but they often impose rigid workflows that make nuanced temporal queries difficult or impossible to express. 
% Even when these tools succeed in generating the desired cohort,  
As a result, analysts and clinicians frequently encounter a trade-off: either wrestle with the steep learning curve of SQL to achieve precision, or accept the limitations of GUI-based builders that restrict expressivity. 
Moreover, the majority of tools are tied to a specific data format, so data extraction pipelines must be rewritten or heavily adapted to accommodate new datasets.
% Neither approach adequately lowers the barrier for non-technical users while preserving the temporal richness required for high-quality ML applications in healthcare. 

% Medical data encompasses a wide array of information, including patient histories, diagnostic images, laboratory results, and treatment plans. 
% This data is crucial for healthcare providers to make informed decisions, ensuring accurate diagnoses and effective treatments. 
% Health research often depends on understanding how medical data unfold over time. 
% For example, a patient may receive a test, then a treatment, and later show a symptom. 
% Writing queries to capture these temporal patterns is hard. 
% Existing tools like SQL or OMOP cohort builders can be powerful, but they are often too complex for non-technical users and can lead to errors or wasted time.

This paper introduces TempoQL, a query language and visual interface designed to simplify authoring, refining, and validating temporal queries on EHR data. It has the following key features:

\begin{enumerate}
    \item \textit{A simple, human-readable yet precise query language.}
TempoQL uses a query language syntax that is designed for readability, so queries can be understood and critiqued by users with a broad range of expertise levels. Unlike natural language interfaces, which can introduce ambiguity and under-specification, TempoQL offers a precise grammar for temporal relations, cohorts, and event sequences. 
% For example, the blood pressure query above could be implemented as \texttt{last \{Mean Blood Pressure; scope = Measurement\} from \#now - 1 hour to \#now every 1 hour}.
% This precision ensures that queries remain reproducible and interpretable, while the readability empowers clinicians to verify that the intent of a query aligns with its execution. By making the query process transparent, TempoQL fosters collaboration between data scientists and healthcare practitioners.

\item \textit{Portability across healthcare data standards.}
% Another central design goal of TempoQL is interoperability. 
TempoQL can operate on diverse data models, including the Observational Medical Outcomes Partnership Common Data Model (OMOP)~\citep{observational_health_data_sciences_and_informatics_book_nodate}, the Medical Event Data Standard (MEDS)~\citep{arnrich_medical_2024}, and the custom formats used by MIMIC-IV and eICU. 
% abstracting away backend differences through a unified query layer. 
This enables researchers to apply queries across multiple datasets and institutions without extensive re-engineering. 
% By separating the logical query from its physical implementation, TempoQL reduces redundant effort and promotes the portability of clinical ML workflows. This feature is particularly valuable in multi-institutional studies where data may be stored in different formats but a consistent analysis framework is required.

\item \textit{Interactive profiling and LLM-assisted authoring.}
Critical to making TempoQL easier to adopt and work with, this work integrates an interactive Python notebook-based interface in which users can inspect intermediate query results and evaluate cohort characteristics. 
% These profiling tools make the data preparation process more transparent, helping users catch errors before they propagate into downstream ML models. 
The interface also incorporates large language model (LLM)–based assistance for writing and refining queries. 
% Users can start from natural language prompts, receive candidate TempoQL queries, and iteratively refine them with system guidance. 
These features reduce the risk of silent errors while accelerating the workflow for both technical and non-technical users.
\end{enumerate}

The goal of this paper is to describe TempoQL's design and explore its potential utility for practitioners in the ML4H community.
We demonstrate TempoQL's key affordances through a lightweight performance evaluation and two example use cases with widely-used data models: identifying cohorts of interest in an OMOP database, and evaluating prediction models for patients with sepsis.

% We begin by reviewing related literature on EHR data querying solutions, then describe the design of the query language and the interactive interface, present case studies demonstrating TempoQL in action, and finally discuss the system's current strengths, limitations, and directions for future efforts.

\section{Related Work}

Several systems have been developed to make it easier to query clinical data, but each has important limits. 
For example, the HL7 Clinical Quality Language (CQL) and the recent ACES cohort extraction system~\citep{xu_aces_2025} help define temporal variables for quality improvement and clinical decision support.
While flexible, versatile, and fairly portable across data models, these tools require technical expertise to author and do not provide interactive capabilities to diagnose and resolve query issues.
Meanwhile, low- or no-code cohort-building systems such as Leaf and COQUITO provide graphical interfaces to create and inspect queries, but they often fail to support rich temporal logic~\citep{dobbins_leaf_2020,krause_supporting_2016}. 
Tools like Atlas enable standardized and reproducible cohort definitions across OMOP datasets~\citep{observational_health_data_sciences_and_informatics_book_nodate}, but these tools are geared towards producing tables for traditional statistical analysis and may be less well-suited for the data pipelines needed to train modern ML methods~\citep{arnrich_medical_2024,Raghu2017}.

% However, these tools rely heavily on structured SQL-like definitions and can be difficult for non-technical users to understand and use.

% Other approaches focus on interoperability standards. FHIRPath and SQL-on-FHIR allow navigation and extraction of data from FHIR resources, but they are not designed to support complex temporal queries for clinical researchers (Brush, 2024). In terms of efficiency, systems like TELII build time-aware indexes to make temporal queries thousands of times faster, but these solutions emphasize backend speed rather than usability or query interpretability (Zhang et al., 2024).
More recently, researchers have explored the use of large language models to generate SQL queries directly from natural language. 
These systems can lower the barrier to database access, but they can be prone to errors in query generation~\citep{shen_text2sql_errors_2025}, especially when tasked with implementing domain-specific temporal logic without extensive prompting or retrieval augmentation ~\citep{gao_text2sql_benchmark_vldb2024}. 
Even the most effective LLMs like GPT-4 perform much worse than humans on large, noisy databases, highlighting the difficulties of schema linking and data noise in real-world settings~\citep{li2023bird}.
% In summary, current LLM-based systems reduce the complexity of query building but lack reliable temporal support and semantic correctness, underscoring the need for structured, domain-aware approaches.
TempoQL's LLM workflow is designed to overcome some of these challenges by generating our intermediate query language representation, which can be inspected and debugged by humans more easily than raw SQL.

\section{Tempo Query Language}

The Tempo Query Language (TempoQL) is designed to support readable yet precise temporal queries that are adaptable to different EHR data standards and can directly feed into ML modeling pipelines.
TempoQL was initially developed as part of a model prototyping system called Tempo~\citep{sivaraman_tempo_2025}, which helped data scientists collaborate with clinicians to improve model specifications.
However, the language used in that work was constrained to use a specific data format (CSV files with schemas designed specially for the Tempo platform), which necessitated considerable preprocessing to import a new dataset.
TempoQL builds on this early implementation with a flexible syntax for \textit{data elements}, combined with a lightweight specification format to bridge between EHR data standards and TempoQL's core data types.
Below we describe TempoQL's approach to extracting data elements from an underlying dataset; then we present the language's principal syntax features, which are an expansion of the language initially developed by~\citeauthor{sivaraman_tempo_2025}.

\subsection{Extracting Data Elements}
\label{sec:data-elements}

In TempoQL, data is considered to be grouped by \textit{trajectories} associated with a unique identifier, typically a patient or visit ID as in existing data standards~\citep{arnrich_medical_2024}.
TempoQL supports data elements with three core types: \textit{Attributes}, which have at most one value per trajectory; \textit{Events}, which have a trajectory ID and a single time point; and \textit{Intervals}, which are similar to Events but take place between two time points.
The goal of data element extraction is to convert a user-provided \textit{data element query} to a set of Attributes, Events, or Intervals, regardless of the format of the underlying data.

\begin{figure}
    \centering
    \includegraphics[width=\linewidth]{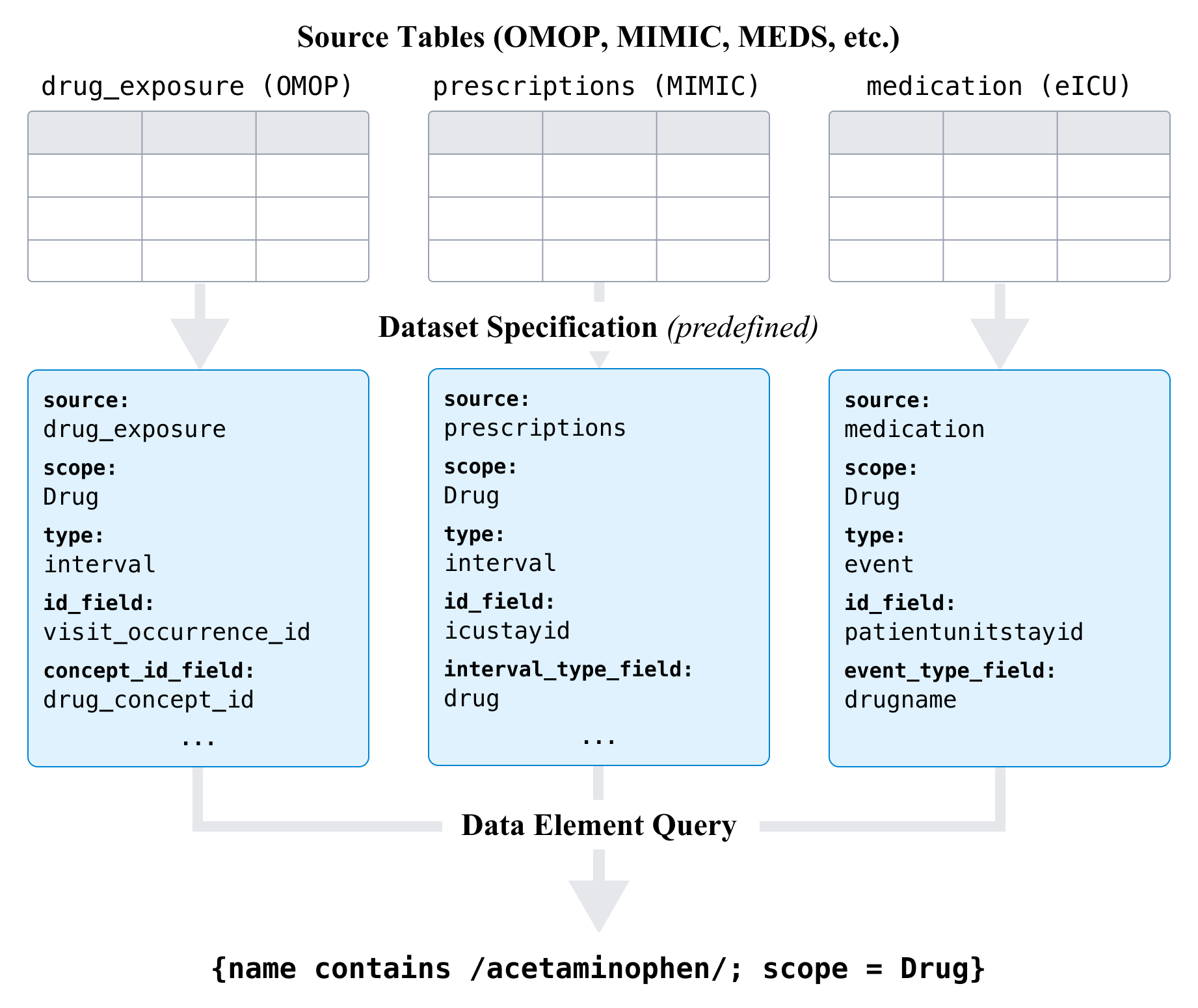}
    \caption{TempoQL data element queries are computed by retrieving data from Source Tables according to a predefined Dataset Specification. This specification allows for querying multiple underlying data models using a consistent query structure.}
    \label{fig:data-element-workflow}
\end{figure}

To accomplish this, TempoQL is initialized with a lightweight \textit{Dataset Specification}: a dictionary containing the available tables and the mapping of source column names to fields required by the core data types.
For example, the specification for the \texttt{drug\_exposure} table on the left of Fig. \ref{fig:data-element-workflow} indicates that the data will be returned as Intervals, the trajectory ID will be the \texttt{visit\_occurrence\_id}, and the type of each interval will be sourced from the \texttt{drug\_concept\_id} column.
% Users can then query for specific interval types by wrapping the name in curly braces, as shown in the bottom left of Fig. \ref{fig:data-element-workflow}.
For Attributes, which are often stored in wide tables where the primary key is the trajectory ID, a list of attribute names mapped to specific columns can be provided.
Dataset specifications for the MIMIC-IV, eICU, MEDS, and OMOP formats are already included in the TempoQL Python library and can be used out-of-the-box (example in Appendix \ref{apd:dataset-spec}), so users of these data models would not need to author a specification at all.
For custom database formats, the specifications can easily be edited and extended.

A frequent and often cumbersome task in EHR data extraction is to combine data across a set of clinical concepts with different identifiers but similar meanings.
For example, there may be several \texttt{drug\_concept\_id} values for very similar drugs; in a traditional query one would have to manually join against an external vocabulary table to identify the relevant concept IDs.
The Dataset Specification can define links to these vocabulary tables, allowing the query engine to perform this join automatically.
As the query at the bottom of Fig. \ref{fig:data-element-workflow} shows, minimal syntax is required to harmonize multiple concepts via the vocabulary table.

The data element query itself comprises a few simple components that can remain consistent across data formats.
Within the curly braces used to signify a data element, one can search for data using five criteria: \texttt{name} (the name of the data element or concept), \texttt{id} (the concept identifier), \texttt{scope} (the source table or vocabulary type to search in, such as ``Medication''), \texttt{type} (the return data type), or \texttt{value} (an alternative column from which to return values).
% \begin{itemize}
%     \item \texttt{name}: Selects by the name of the attribute, event, or interval, or the concept name.
%     \item \texttt{id}: Selects based on the concept identifier.
%     \item \texttt{scope}: Limits the search to a particular ``scope,'' which usually corresponds to a source table or vocabulary type. For instance, the specification in Fig. \ref{fig:data-element-workflow} defines ``Medication'' or ``Measurement'' as scopes. Data element names must be unique within a scope.
%     \item \texttt{type}: The core data type that should be returned, such as ``attribute,'' ``event,'' or ``interval.''
%     \item \texttt{value}: Selects a column in the source data other than the default value field. This can be used to retrieve secondary data associated with an event or interval, such as the unit of measurement for a drug dosage.
% \end{itemize}
The simplest data element query is an exact name match wrapped in curly braces, e.g. \texttt{\{Gender\}}.
More complex data elements can combine multiple criteria and define concept sets, such as \texttt{\{name in ("Respiratory Rate", "Resp Rate")\}} or \texttt{\{scope = Observation; name contains /(breath|resp\textbackslash w*) rate/i\}}.

A key insight behind TempoQL's portability is that downstream query processing and aggregation can use the same Pandas-based implementations as long as the data element queries return a core data type (Attributes, Events, or Intervals). 
We explored rendering entire TempoQL queries into SQL for fully database-side processing, as is done by some existing tools~\citep{schuemie_health-analytics_2024}, but we found that in our case the benefits of database-agnostic query implementations outweighed the potential performance pitfalls of aggregating data locally.
TempoQL uses SQLAlchemy Core\footnote{\url{https://www.sqlalchemy.org}}, a database-agnostic Python API for building SQL queries, to implement most currently-supported formats (OMOP, MIMIC-IV, eICU).
Data elements for MEDS, meanwhile, use a different implementation directly based on Pandas because MEDS datasets typically combine multiple scopes into sharded Parquet files.
Regardless of how data elements are retrieved, the syntax and underlying implementation for further processing is identical, as described below.

\subsection{Readable, Precise Temporal Transformations}

Data elements can be transformed and aggregated in TempoQL using syntax that is both inspired by SQL and geared towards natural-language-like expressions.
Basic arithmetic and logical expressions are very similar to SQL, and can automatically handle broadcasting between elements of different sizes, such as Attributes and Events.
For example, the following query extracts and formats body temperature measurements by checking for values that were likely mistakenly entered in Fahrenheit and converting them to Celsius, then filtering out outliers:
\begin{quote}
\small
\begin{verbatim}
(case
  when temp > 45 then (temp - 32) * 5 / 9
  else temp
end 
where #value between 20 and 50) 
with temp as 
  {Body temperature; scope = LOINC}
\end{verbatim}
\end{quote}

\begin{figure}[t]
    \centering
    \includegraphics[width=\linewidth]{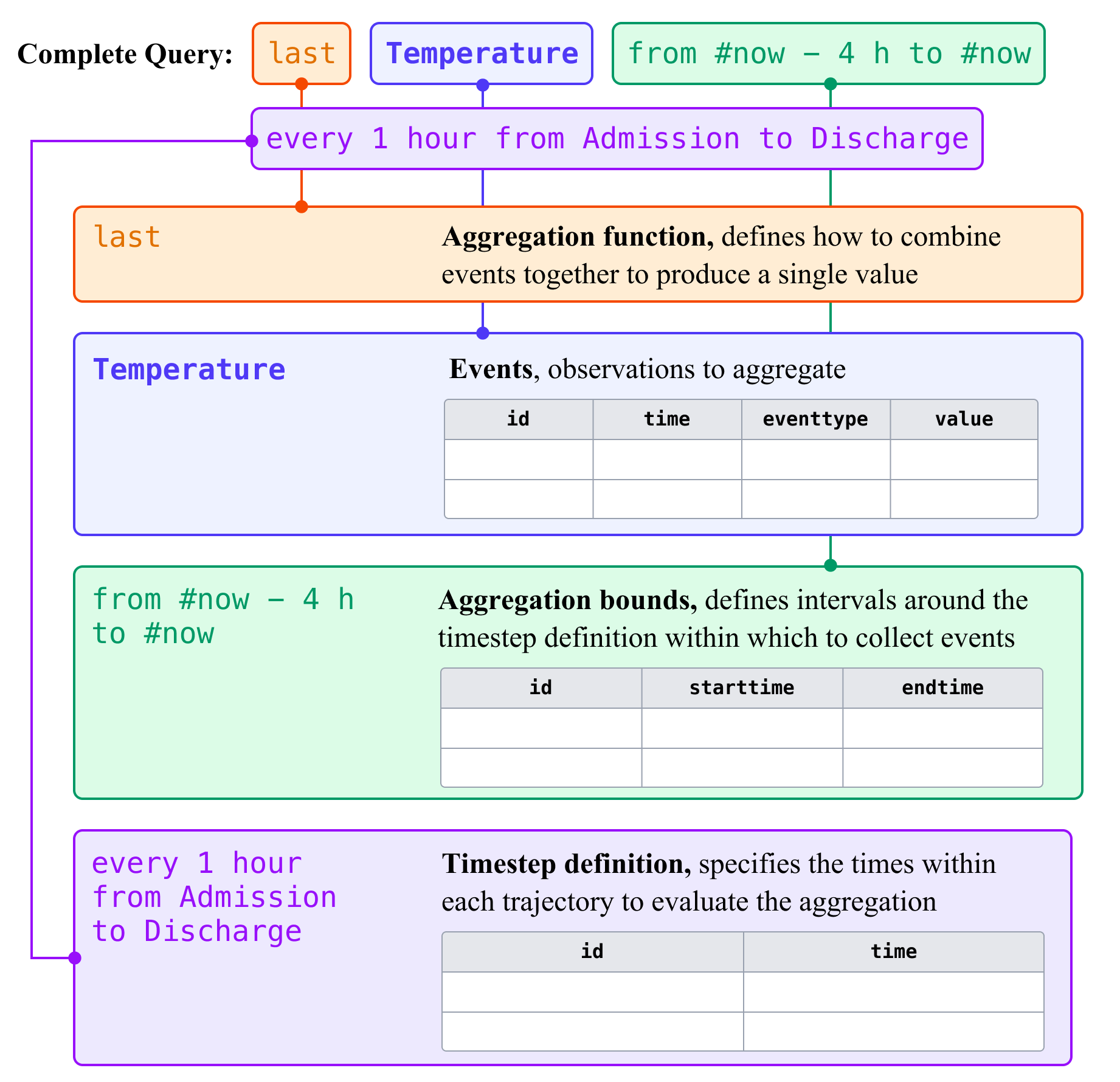}
    \caption{TempoQL's temporal aggregation syntax consists of an aggregation function, the Events or Intervals to be aggregated, the aggregation bounds at each timestep, and the timestep definition. The query above produces a rolling average of the patient's temperature for each hour in their stay.}
    \label{fig:aggregations}
\end{figure}

Next, \textit{temporal aggregations} are the main computational component of TempoQL's query engine, and they provide the most support towards creating well-structured tables for machine learning.
As described in the first version of Tempo \citep{sivaraman_tempo_2025} and visually depicted in Fig. \ref{fig:aggregations}, aggregations in TempoQL are specified by three or four components. First, the aggregation function (such as \texttt{mean}, \texttt{first}/\texttt{last}, \texttt{exists}, etc.) defines how to combine values from the second component, the Events or Intervals being aggregated. 
Third, the aggregation bounds specify time windows within which to collect matching events or intervals.
Finally, a timestep definition specifies index time points at which to compute the aggregation, which can be regular intervals (\texttt{every 24 hours}), times that an event occurs (\texttt{at every start(\{Visit\})}), or a single value for each trajectory.
Together, these elements provide a conceptually intuitive way to define, edit and iteratively improve aggregation queries.

TempoQL provides additional built-in syntax elements for time-series preprocessing and data cleaning.
For example, the \texttt{where} keyword allows for filtering out unwanted values, while the \texttt{impute} keyword replaces missing values with results of another expression.
Values can be carried forward into rows with missing values using the \texttt{carry} keyword.
To create categorical variables, numerical data can be discretized with the \texttt{cut} command.
Further functionality is implemented using Python-like function syntax, such as \texttt{extract()} to perform pattern matching with regular expressions and \texttt{intervals()} to combine two sets of Events into an Intervals object using a time-based join.
This syntax pattern allows TempoQL to be more easily extended to future use cases.

\section{TempoQL Interface}
\label{sec:interface}

Even with TempoQL’s more readable query language design, some practical difficulties still remain for reliably extracting EHR data, as we learned from the initial version of Tempo~\citep{sivaraman_tempo_2025}. 
Regardless of which query language is used, users often lack the ability to easily see “what’s in the data,” making it difficult to formulate the right query to extract data they want. 
Moreover, after a query runs, inspecting intermediate results and diagnosing errors is cumbersome. 
Finally, users may still face difficulties learning a new query language and using it to write potentially complex queries. 
To overcome those challenges, we designed and developed the TempoQL interface, which runs within an interactive Python environment as shown in Fig. \ref{fig:ui-overview}.

\paragraph{Browsing Data Elements.}
\label{sec:sub-data-elements}
The \textit{Data Elements} section (Fig. \ref{fig:data-element}) can help users discover what information is available in the dataset and start generating their queries. 
When the interface is initialized, it automatically scans the connected database and compiles a catalog of available data elements. 
% For each element, the system provides metadata such as the table it belongs to, its time-series type (e.g., event, interval), and the total number of occurrences in the dataset. 
Users can then search for specific concepts, browse through standardized vocabularies, and select multiple elements to construct queries. 
% This combination of structured metadata, search functionality, and multi-selection streamlines the process of query generation, reducing reliance on prior knowledge of database schemas and enabling more efficient exploration of the dataset.

\paragraph{Query Editor.}
% The central feature of TempoQL’s interface is its query construction and execution capability. 
% Unlike traditional approaches that require programming skills, the system provides a graphical environment where users can write and run queries directly. 
The TempoQL interface's editing features are designed to enable rapid and intuitive experimentation while improving reproducibility.
The Query Editor includes auto-completion for keywords, syntax highlighting, and the ability to retrieve prior queries from history. 
In addition, we developed a convenient system to persist interactive edits: users can specify a JSON file-path at which to read and write a collection of queries.
This file can then be versioned and shared among collaborators as well as directly passed to the TempoQL query engine to extract data in non-interactive Python environments.
% In addition, the interface provides mechanisms to import and export queries, enabling seamless communication among collaborators and convenient reuse across different projects. 

\paragraph{Profiling and Debugging Query Results.}
To help users understand the data returned by a query, TempoQL introduces a dedicated Query Results section in the right sidebar. 
Once a query has been executed successfully, the system presents intuitive visual summaries that allow users to quickly assess the extracted data. 
Inspired by prior systems for data profiling~\citep{epperson_dead_2023}, the sidebar includes high-level summaries of counts (by patient trajectory as well as within each trajectory), missingness, and the distribution of values. 
Furthermore, when a query contains subcomponents (e.g., data element queries, variable references, or aggregation expressions), the results of these ``subqueries'' are displayed in the sidebar (see Fig. \ref{fig:subqueries}).
For subqueries that run SQL on the database, the code rendered by SQLAlchemy is provided to facilitate understanding of what the TempoQL query is doing.
By surfacing these descriptive insights immediately, the Query Results section helps users validate whether the query produced the intended cohort, identify anomalies, and iteratively refine their queries. 
The raw data can also be accessed within the same Python notebook if additional analyses on the results are desired.
% This tight feedback loop deepens users’ understanding of the data and reduces the risk of silent errors propagating into downstream analyses.

\paragraph{LLM-Assisted Query Authoring.}
% \begin{figure*}
%     \centering
%     \includegraphics[width=\linewidth]{images/ai-workflow.png}
%     \caption{Caption}
%     \label{fig:ai-workflow}
% \end{figure*}
To help reduce the learning curve for people to use TempoQL, we integrate large language models (LLMs) into the workflow to help users write, fix and explain queries. 
The design of this ``AI Assistant'' is inspired by prior work on using language models as interfaces to structured databases~\citep{li_nalir_2014,ziletti_retrieval_2024}. %, we implement an ``AI Assistant'' that enables TempoQL system to apply large language models in assisting the process of querying healthcare data. 

% In TempoQL system, the LLM acts as an assistant throughout the query lifecycle. 
Users can begin by prompting the LLM to generate or update a query using natural language, which the LLM translates into a candidate TempoQL query. 
The user can then edit and run this query directly in the interface. 
The user can also request an explanation of the query with a single click. 
% LLM can further support interpretation by providing plain-language explanations of what the query extracts and offering debugging suggestions when errors occur. 
This pipeline positions the LLM as an interactive collaborator that does not perform queries itself, but rather helps users create, refine, and understand their analysis process. 
% By integrating the LLM in this way, TempoQL substantially lowers the barrier for non-technical users, such as clinicians, to engage directly with complex healthcare data. 

Currently, we use Gemini 2.5 Pro\footnote{\url{https://cloud.google.com/vertex-ai/generative-ai/docs/models/gemini/2-5-pro}} for few-shot query generation and explanation, using a prompt that describes TempoQL's syntax in detail and gives several examples of correct queries (see Appendix \ref{apd:llm-prompts}). 
Other LLMs can easily be substituted in the future to accommodate security constraints.
We also consider ways to give the LLM context about the dataset while preserving privacy. 
% To solve this, we build our prompt consisting of both TempoQL and dataset information. The first part includes TempoQL syntax, operator/aggregation semantics, and valid constructs which acts as a strong prior that narrows the output space to well‑formed queries and concise explanations. 
To avoid exposing patient data directly to the LLM, we instead allow the model to use function-calling to request information about Data Elements relevant to the user's question, including scopes and concept IDs and names. More implementation details of function calling, including the scope of data accessible to and hidden from the LLM, are provided in Appendix \ref{apd:function-call}.
% To improve the robustness of the system, the system incorporates the function calling mechanism that converts natural language to structured queries. In addition, error-handling and explanation modules help prevent unreliable system behavior and provide users clear feedback, guiding towards efficient corrections. 
% These design choices contribute to better performance and parsimonious data sharing when integrating LLMs into the query workflow.

\section{Performance Evaluation}
To evaluate the speed, accuracy, and expressiveness of TempoQL and its LLM-assisted authoring workflow, we designed 12 basic types of queries that users may commonly want to perform (see Appendix \ref{apd:perf-eval} for all queries). 
We then compared the results of these queries to a conventional workflow using BigQuery SQL by their execution time, query generation accuracy, and conciseness. 

\begin{figure}[t]
    \centering
    \includegraphics[width=\linewidth]{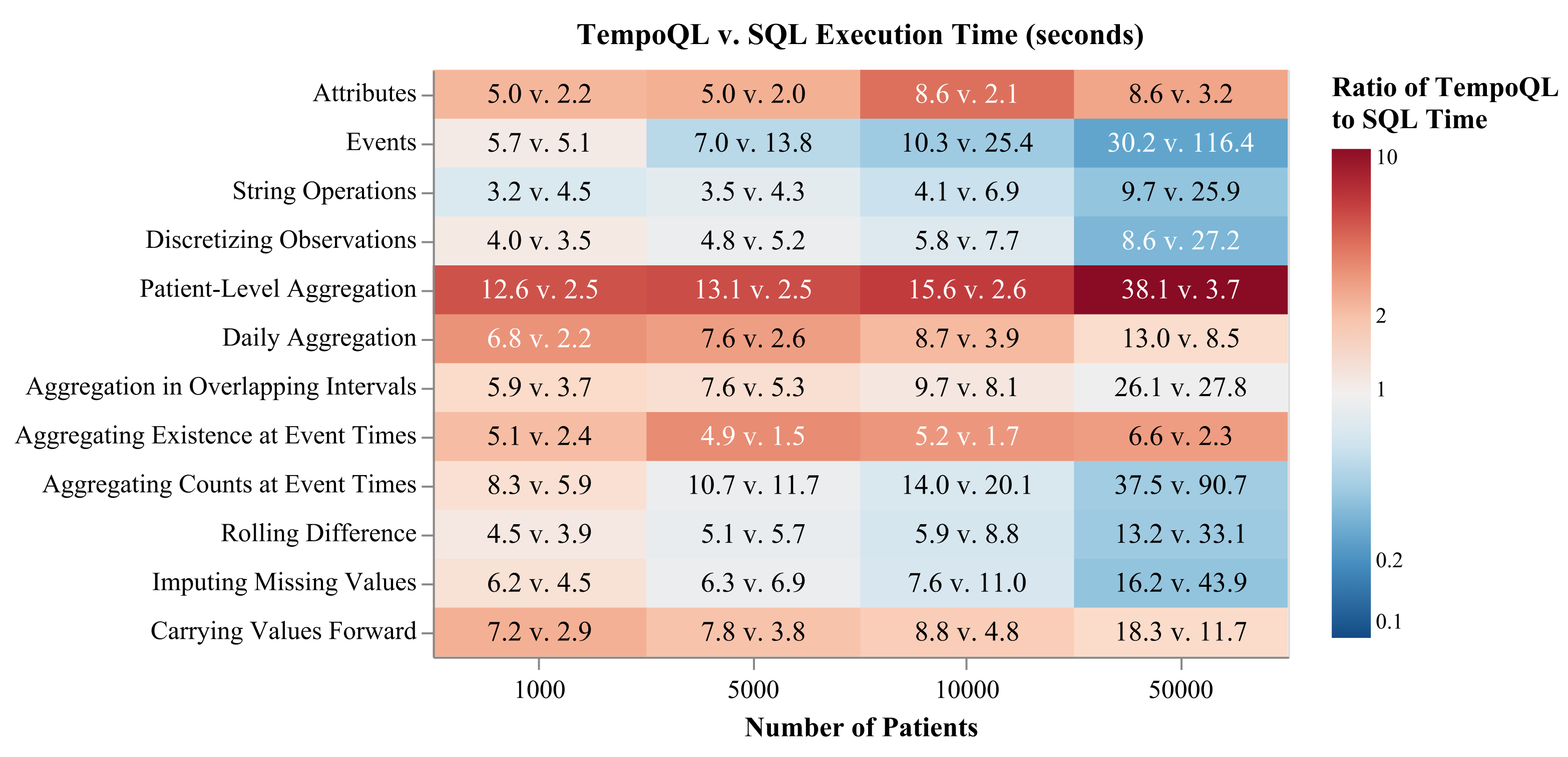}
    \caption{Difference in execution time between TempoQL and BigQuery SQL across dataset sizes, averaged over three trials.}
    \label{fig:performance-eval}
\end{figure}

\paragraph{TempoQL Query Execution Time}
For each query type, we created equivalent TempoQL and BigQuery SQL queries that we could execute on the MIMIC-IV dataset. 
We then ran these queries on a single machine over subsets of 1K, 5K, 10K, and 50K ICU stays with three replicates each (result tables range in length from 300 to 4M rows). 
As the results in Fig. \ref{fig:performance-eval} show, TempoQL achieves comparable performance to BigQuery SQL across most of the query types, and exhibits better scalability than BigQuery SQL for some of the queries. 
This advantage may be because TempoQL’s data element retrievals more often align with the prebuilt indexes for the dataset, and in-memory aggregations can avoid data transfers needed to compare values. 
On the query types where TempoQL is slower than SQL (e.g. Patient-Level Aggregation), we hypothesize the main bottleneck is the data download: TempoQL must download all events for every patient and aggregate them locally, while SQL aggregates these events database-side and sends only the relatively compact result.
Nevertheless, this bottleneck does not appear to affect performance for most of the 12 query types, indicating that TempoQL would introduce minimal computational overhead relative to a traditional SQL-based workflow.

\paragraph{LLM Query Generation Accuracy}
For each query type, we also created natural-language prompts describing the desired results. 
We created a few-shot prompting setup for MIMIC-IV on BigQuery that was very similar to the one used for TempoQL, providing all table context needed to generate SQL code for each query type.
From the results in Table \ref{fig:LLM-data-results}, we see that Gemini’s generated TempoQL is about half as likely to throw an error than Gemini-generated SQL, and it is 2.5 times more likely to return the correct result. 
% The high correctness rate and low incidence of errors indicate that LLM-assisted TempoQL query generation is generally robust. 
Most errors in the generated SQL arose from type errors, despite having access to database schema information in the prompt, and unnecessary filters and joins that changed the result correctness.
We hypothesize that TempoQL's simplicity helps mitigate these types of errors.

\paragraph{Query Conciseness}
Comparing the TempoQL and SQL queries in Appendix \ref{apd:perf-eval}, TempoQL queries demonstrate substantially greater compactness than their SQL equivalents, containing on average 5.9 times fewer code tokens. 
This reduction in syntactic complexity not only improves readability but also reduces the likelihood of subtle, hard-to-detect implementation errors.

\label{sec:perf-eval}

\begin{table}[t]
\centering
\small
\begin{tabular}{r c c c}
\toprule
\textbf{Method} & \textbf{Correct} & \textbf{Incorrect} & \textbf{Invalid} \\ \hline
\textbf{TempoQL} & 81.5\% & 1.7\% & 16.8\% \\
\textbf{BigQuery SQL} & 31.9\% & 35.3\% & 32.8\% \\ 
\bottomrule
\end{tabular}
\caption{Accuracy of Gemini 2.5 Pro in generating TempoQL and BigQuery SQL from natural language prompts. Rates are computed over 10 calls to Gemini for 12 query categories.}
\label{fig:LLM-data-results}
\end{table}

\section{Use Cases}
\label{sec:use-cases}

We provide two example use cases to demonstrate TempoQL's expressiveness, portability, and ease of use on real-world problems.
Since TempoQL is a tool to support ML4H research, the goal of these use cases is to provide readers with a sense of the opportunities opened by the system as well as areas for further development.

\subsection{Exploring Patient Cohorts for Adverse Events}

Drug-related problems such as potentially inappropriate medications (PIMs) and adverse drug reactions (ADRs) are of increasing concern as the global population ages~\citep{zazzara_adverse_2021,tian_prevalence_2023}.
% These issues may stem from not only clinical factors but also communication issues, difficulties with medication awareness and literacy, and poor care coordination. %are some of the common issues that underly the high rate of drug related problems experienced by older adult patients. 
% Particularly risky time points include transitions of care such as discharge from a hospital to home. 
% Medication changes are often not communicated effectively to the patient and informal caregivers. 
% Patients struggle to untangle look-alike pills and confusing administration instructions. 
% Making matters worse, many non-medical factors influence health outcomes such as lack of stable housing, poor access to transportation, and social isolation. 
In order to develop predictive models that could be used in clinical interventions to improve  medication safety, we previously compiled a dataset of nearly 600,000 patients across hospitals, skilled nursing facilities, and home health care.
% The dataset is person-centered and crosses multiple care settings (). 
The OMOP Common Data Model was used to represent the data, covering conditions, medication exposures, procedures, lab measurements, demographics, observations, and family histories.
Using TempoQL, we can quickly explore possible drug-related issues in this dataset by constructing precise, readable cohort definitions.
% We have developed separate machine learning models to predict falls, gastrointestinal bleeds, and stroke within 90 days of discharge from a hospital to the community outpatient setting.

For one example, let us investigate a possible association between semaglutide and acute kidney injury (AKI), which has previously been described in case studies~\citep{leehey_acute_2021}.
Using TempoQL's interface, we first search for drugs by the names ``semaglutide'' and its commercial name ``Ozempic'' in the \texttt{Drug} scope, which results in two matching concepts.
Running a data element query for these concepts results in 129 occurrences in the database.
Next, we perform the same process for kidney injury, using a SNOMED concept that has about 59,000 occurrences in the data.
Putting the two together, we can assign the two queries' values to the variables \texttt{semaglutide\_rx} and \texttt{aki\_outcome} and write an aggregation expression to find occurrences of one before the other: \texttt{exists aki\_outcome after semaglutide\_rx}.
However, this initial query may be too broad, so we can make it more specific by restricting the time window of AKI to 90 days after the prescription and excluding patients with AKI history:
\begin{quote}
\small
\begin{verbatim}
(exists aki_outcome 
  from first_rx to first_rx + 90 days) 
where (not exists aki_outcome before first_rx)
with first_rx as (
  first starttime(semaglutide_rx) 
  from #mintime to #maxtime
)
\end{verbatim}
\end{quote}
This query results in only three matching occurrences, perhaps because our data only ran until 2018, when semaglutide was just starting to be widely prescribed.
Next, we could manually inspect these cases to understand them in more detail, or broaden our queries to find a larger cohort.
TempoQL helped us quickly find relevant patient cases matching the conditions of interest, while keeping a precise and readable record of our exploration process.

\subsection{Evaluating Sepsis Outcome Prediction Models Across Datasets}
\label{sec:use-case-sepsis}

Sepsis is a leading cause of mortality in hospitals and a focal point for efforts to improve outcomes for patients in the intensive care unit (ICU)~\citep{SocietyofCriticalCareMedicine2021}.
Deep learning techniques can help drive these efforts by learning latent patterns from large cohorts of sepsis patients~\citep{Killian2020}, but it can be difficult to adapt these models to new settings.
% Moreover, commonly-used databases such as MIMIC may not contain enough diversity in decisions to learn accurate relationships between treatments and outcomes~\citep{park_how_2024}.
TempoQL provides an opportunity to explore the generalizability of predictive models for sepsis across multiple datasets.

We derive datasets from three publicly-available databases: MIMIC-IV~\citep{johnson_mimic-iv_2023}, the eICU Collaborative Research Database~\citep{pollard2018eicu}, and EHRSHOT-MEDS~\citep{wornow2023ehrshot}.
Since all three databases are provided in different formats, an entirely different pipeline for each database would typically be required.
With TempoQL, we can use a single Python script to perform data extraction on any of the three datasets, using JSON files to store queries adapted to each format.
As shown in Fig. \ref{fig:sepsis-model-results}(a) and described in more detail in Appendix \ref{apd:second}, we create the datasets in three stages: selecting the patient cohort, extracting dataset-specific raw data, and aggregating observations to create model features.
The first stage determines which patients will be included in the downstream dataset; % in this case, patients over the age of 18 who spent at least 4 hours in the ICU and met sepsis criteria or received a relevant diagnosis code.
then, we create dataset-specific variants for each model feature.
These queries are generally short and similar across data formats but may require minor adaptation because observations are frequently stored under different scopes or concepts.
TempoQL's AI Assistant feature can help accelerate this adaptation by generating alternative versions of the queries, such as by converting an ICD-10-based diagnosis query from MIMIC to use SNOMED codes in EHRSHOT.
Finally, we use a single shared set of queries to aggregate the raw variables at four-hour intervals and produce the model input features.

TempoQL can help us identify potential issues with these datasets during and after model training.
For example, after training a model on MIMIC-IV to predict Time to Discharge (Fig. \ref{fig:sepsis-model-results}b), we observe that the model generalizes fairly well for eICU but drops significantly in performance for EHRSHOT.
TempoQL's Query Results view for the outcome variable reveals higher average discharge times in EHRSHOT (around 150 hours compared to 112 hours).
Since the earlier stages of data extraction are also implemented in TempoQL, we can browse them to find that the Discharge query in EHRSHOT relies on the existence of a SNOMED discharge event, while the other two datasets directly surface visit intervals.
For the purpose of constructing trajectories, our queries had imputed the discharge event times with a default time-point when missing, but the imputed values would not be appropriate for prediction.
Therefore, we can add a \texttt{where} clause to our Time to Discharge variable to filter out missing Discharge timestamps, preventing the model from being evaluated on these instances.
If we had used separate preprocessing code for all three datasets, it would have been difficult to pinpoint which differences resulted in the apparent model performance drop.
Using TempoQL resulted in a simpler, more consistent pipeline, making it easier to interpret and debug the downstream results.
% Using these datasets, we train autoencoder models on the MIMIC-IV dataset using simple dense networks and self-attention layers (transformer).
% We evaluate the models using simple linear probes to perform four downstream prediction tasks on all three datasets: discharge outcomes of mortality, time to ICU discharge, use of vasopressors in 12 hours, and change in SOFA score (a measure of disease severity) in 12 hours.
% As shown in Fig. \ref{fig:sepsis-model-results}, transformer-based methods always outperform dense networks but the relative performance across datasets differs based on the predictive task.
% For example, we find that predictive performance on EHRSHOT is worse than the other datasets for Mortality and Hours to Discharge, but better for Vasopressor Need and Change in SOFA Score.
% These discrepancies could be due to differing availability of the input features and outcomes, which we can investigate and resolve using TempoQL's query profiling tools.

\begin{figure}[t]
    \centering
    \includegraphics[width=\linewidth]{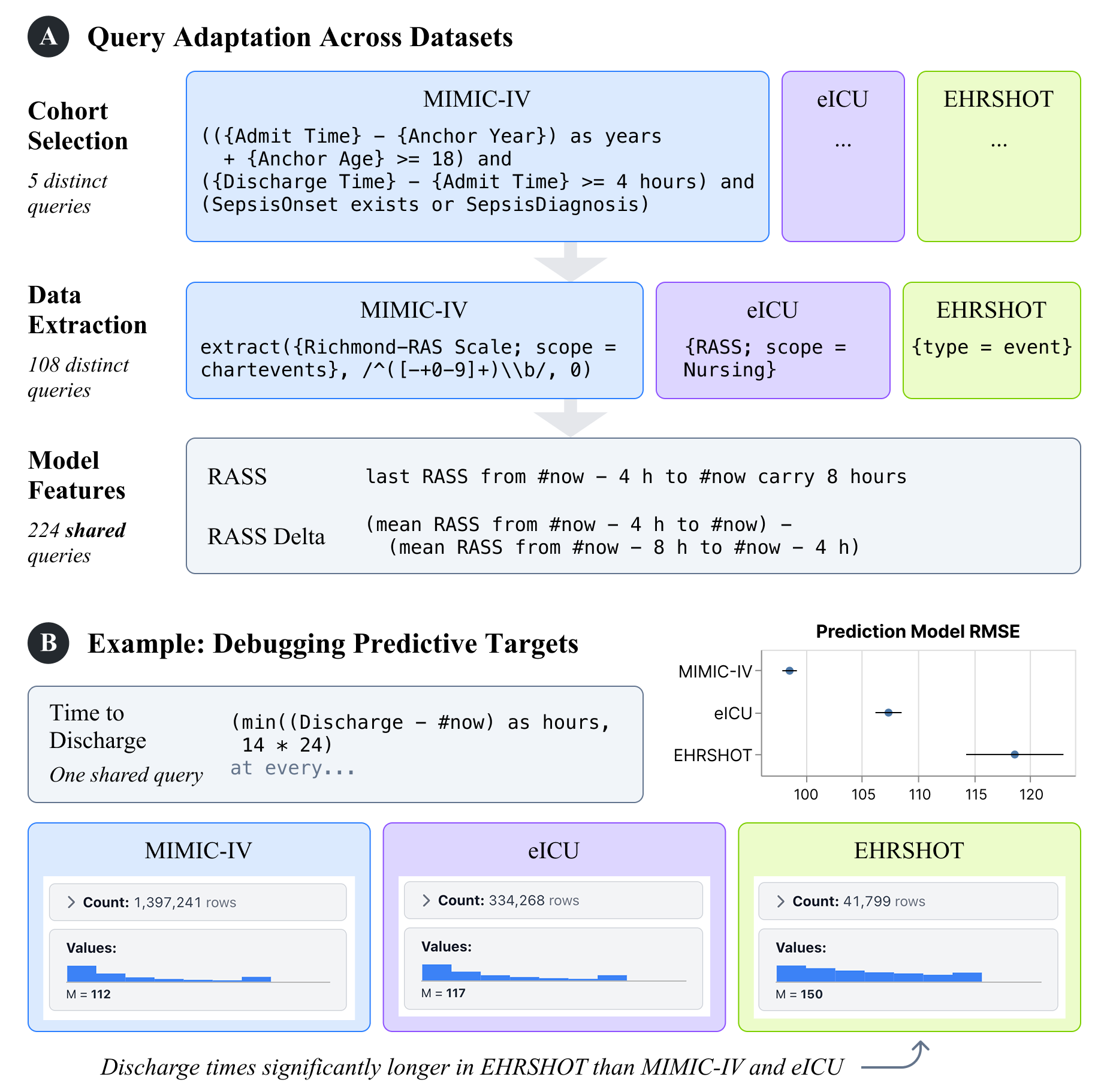}
    \caption{(a) Process for generating parallel datasets for sepsis outcome prediction from three databases with different formats; (b) example of debugging a performance drop in the EHRSHOT dataset for a model predicting time to discharge.}
    \label{fig:sepsis-model-results}
\end{figure}

\section{Discussion}

% \begin{enumerate}
    % \item \textbf{Brief summary of work}\\
In this paper we introduced TempoQL, a query system designed to make electronic health record (EHR) data more accessible, precise, and portable for machine learning applications. 
TempoQL combines a human-readable query language with an interactive data profiling and authoring interface with LLM-assisted support. 
% The language provides clinicians and researchers with a way to write and critique temporal queries without relying on the technical complexity of SQL, while still maintaining the precision required for reproducible analysis. 
The lightweight Dataset Specification enables portability across multiple data standards and institutions. 
% The system also introduces a jupyter notebook-based interface that help users explore available data elements, inspect query outputs, and refine cohort definitions through visualization and profiling tools. 
% We also integrate LLM-based assistance for query generation, explanation, and debugging. 
Together, these components allow TempoQL to bridge the flexibility of query languages and the accessibility of graphical tools, helping users across technical expertise levels develop precise, understandable data extraction processes.
% \item \textbf{Applications of the work: what's possible now that we have TempoQL. Some ideas could be using TempoQL to apply existing ML models to new datasets; assessing differences between the expressivity of different data formats, and debugging conversions between data standards.}

One important use case for TempoQL is in exploring model generalizability and robustness (shown in Sec. \ref{sec:use-case-sepsis}). 
By decoupling query logic from specific database schemas, TempoQL makes it straightforward to apply existing machine learning models to new datasets without rewriting complex extraction pipelines. 
This portability is valuable in doing research across institutions, where models often fail to transfer due to incompatible data representations~\citep{arnrich_medical_2024}. 
Researchers could also use TempoQL to compare the expressivity of different data standards by executing the same temporal queries across different data formats. %, researchers can directly assess which standards support richer clinical concepts and where limitations arise. 
% Such analyses can inform both methodological choices and improve the interoperability of healthcare standards.

Future improvements to TempoQL could further enhance its utility for the ML4H community. 
One promising direction is to integrate TempoQL into graphical cohort builders as an intermediate layer between no-code interfaces and SQL-based systems, allowing users to benefit from the ease of graphical tools while retaining the precision and portability of TempoQL queries. 
Future work could also extend TempoQL to incorporate structured querying on unstructured clinical notes, making it more applicable to tasks where narrative documentation provides critical signals. 
Recent advances in information extraction~\citep{huang_chatgpt_2024,ma_chase_2025} highlight promising directions for bridging this gap. 
% Another opportunity is to use TempoQL public querying and exploration of open-source datasets such as MIMIC-IV and eICU. By providing a consistent, interpretable query layer, TempoQL could serve as a shared platform for education, benchmarking, and reproducible research, where researchers and clinicians collaboratively explore datasets, exchange validated queries, and accelerate the development of robust machine learning models. Beyond these directions, future extensions could expand TempoQL’s scope beyond structured data to include diverse modalities and larger-scale deployments, further increasing its impact in healthcare machine learning.

% \item \textbf{Limitations (and why those limitations are minor). 
% TempoQL currently only works on structured EHR data; it could be helpful to add features that let users structure free-text data using TempoQL. Also, the queries currently run in Pandas, not database-side. Although we have accelerated query computation using Numba, its performance could be improved further.

While TempoQL makes important advances towards making temporal cohort queries more readable and portable, it currently has some limitations that could be addressed in future work. 
% The system is designed primarily for structured EHR data and does not yet natively support unstructured modalities such as clinical notes, radiology reports, or free-text fields.  
One major constraint is that query execution is implemented using Pandas rather than taking place within a database. 
This design simplifies integration across diverse data formats, but it could reduce performance compared to database-native execution in some types of queries as shown in Sec. \ref{sec:perf-eval}. 
Although core computations have been accelerated with Numba, further gains could be achieved by combining database-side processing with local aggregation (e.g., ~\cite{huang_telii_2024}).
% The underlying SQL queries could also be further optimized to the table structure and indexes in each data model, and 
Local computation could also be heavily parallelized since the results are independent per-trajectory.

This paper is a first step towards understanding TempoQL's potential to support querying on EHR datasets, and it should be interpreted with some caveats.
It remains necessary to evaluate TempoQL's usability with ML practitioners through a user study.
Furthermore, our evaluations did not benchmark TempoQL against systems that have been designed specifically for EHR data, such as Atlas or CQL.
However, these query systems often run on SQL under-the-hood, so the differences in execution time are likely consistent with those observed in our study.
Many ML in healthcare tasks also still rely on SQL because of the data model and infrastructure constraints of other EHR-specific tools; TempoQL offers an advantage in these workflows because of its portability.
    % \item \textbf{Future Work (how can we improve TempoQL and create future tools). Some ideas include integrating TempoQL into graphical cohort builders as an intermediate layer between no-code and SQL-based environments; using TempoQL to support public querying and exploration on open-source datasets.}

    % \item \textbf{Summary}\\
While algorithmic innovations have driven progress in ML for healthcare, building and scaling these techniques requires intuitive, shared conceptual representations of the data they are built upon. 
This work highlights the potential of query languages augmented by interactive, LLM-in-the-loop workflows to make EHR data more accessible and ready for modern ML pipelines. 
 % can lower barriers to authoring and understanding analyses on clinical data, enabling more transparent and collaborative healthcare research. 
TempoQL is open-source and available for ML4H researchers to use, extend, and customize for their own data models and applications.
% By making complex data more usable, we move closer to a future where clinical insight and computational methods can be combined more seamlessly to improve patient outcomes.
    
% \end{enumerate}

\acks{We would like to thank Octavius Tan, Maxwell Huang, and Ryan Ng for contributions to the TempoQL codebase, and Xiaotong Li and Dominik Moritz for critical feedback on the language and interface design. The authors were supported by a Carnegie Mellon University Summer Undergraduate Research Fellowship and an NSF Graduate Research Fellowship (DGE2140739).}

\bibliography{jmlr-sample}

\clearpage

\appendix

\section{Additional System Details}

\subsection{Interface Features}
\begin{figure}[t]
    \centering
    \includegraphics[width=\linewidth]{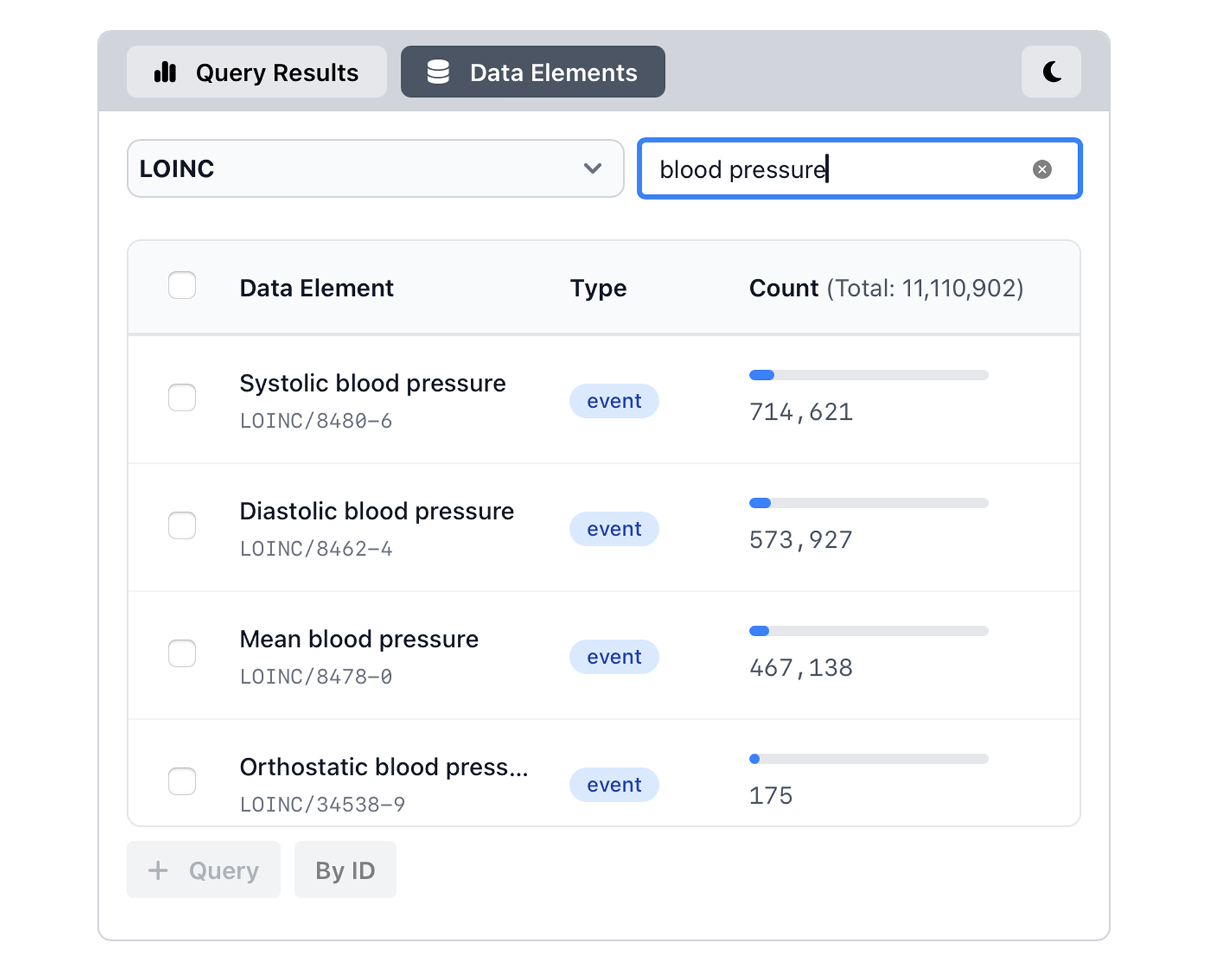}
    \caption{Data Elements Interface Section}
    \label{fig:data-element}
\end{figure}

Fig. \ref{fig:data-element} shows the Data Elements interface described in Section \ref{sec:interface} (\textbf{Browsing Data Elements}). It allows users to search and select standardized medical concepts (e.g., blood pressure) across vocabularies such as LOINC, displaying element types, codes, and data counts for easy query construction. The user can also directly copy the corresponding query by selecting the data element here.

\begin{figure}[t]
    \centering
    \includegraphics[width=\linewidth]{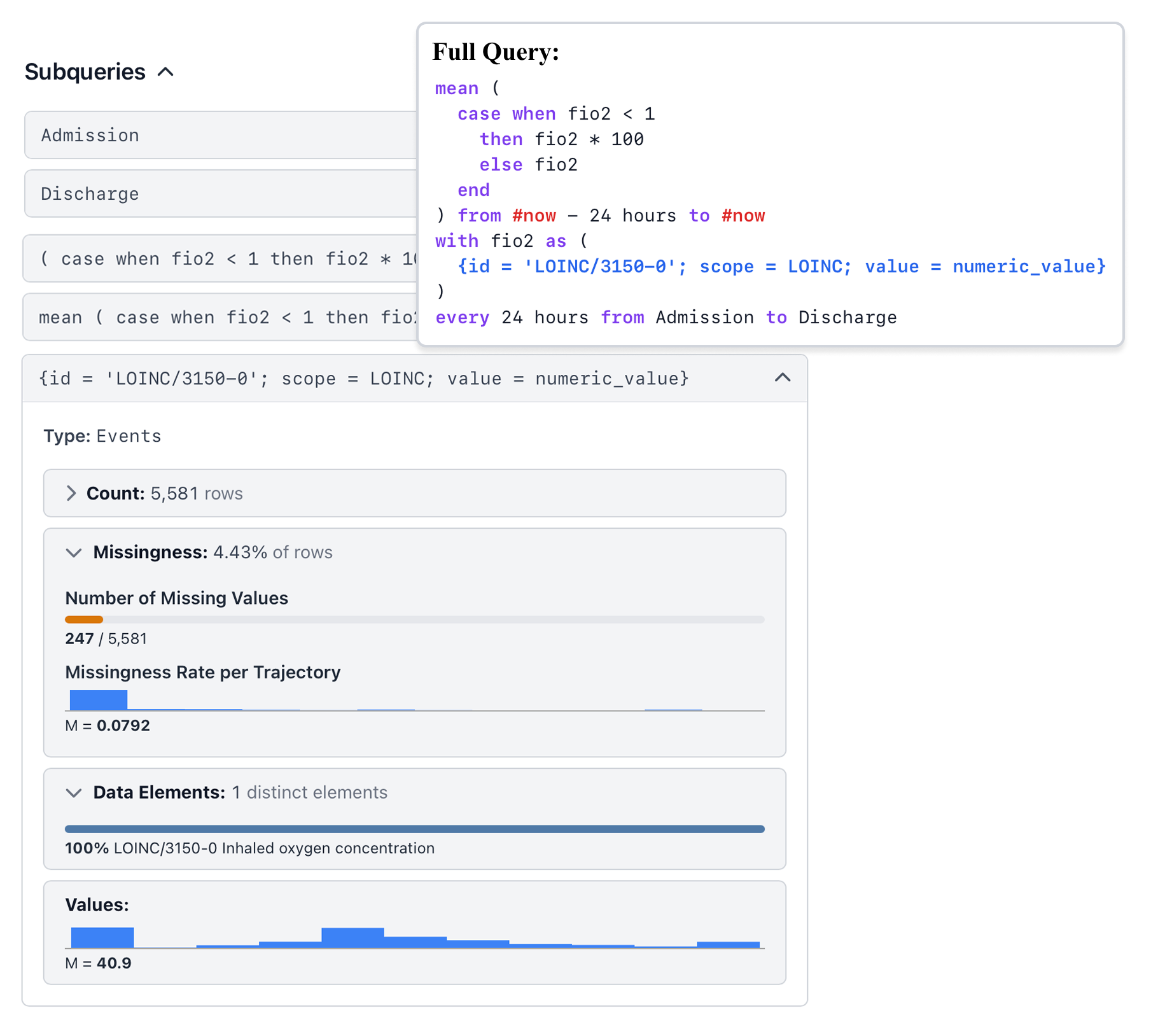}
    \caption{Subqueries Interface Section}
    \label{fig:subqueries}
\end{figure}

Fig. \ref{fig:subqueries} shows the Subqueries Interface described in Section \ref{sec:interface} (\textbf{Profiling and Debugging Query Results}). The Subqueries section lists each valid subquery from the full TempoQL query. For each subquery, the results display the data type, row count, missingness statistics, and corresponding data elements, helping users understand the data underlying each query component.

\subsection{LLM Function Calling Details}
\label{apd:function-call}

In our current implementation, the LLM never accesses raw or patient-level data. Instead, it interacts only through a restricted function-calling interface (\texttt{search\_concepts}), which returns metadata such as concept names, IDs, and scopes. This ensures that the model “sees” only schema-level information needed to construct TempoQL queries, while data access remains entirely local to the user’s environment. The function calling schema is
\begin{small}
\begin{verbatim}
search_concepts_function = {
  "name": "search_concepts",
  "description": "Search for concepts that 
  match a given query. 
  Returns a list of up to 100 concept names 
  that match the query.",
  "parameters": {
    "type": "object",
    "properties": {
      "query": { ... },
      "scope": { ... },
    },
    "required": ["query"],
    ...
  }
}
\end{verbatim}
\end{small}

\begin{figure*}[t]
    \centering
    \includegraphics[width=\linewidth]{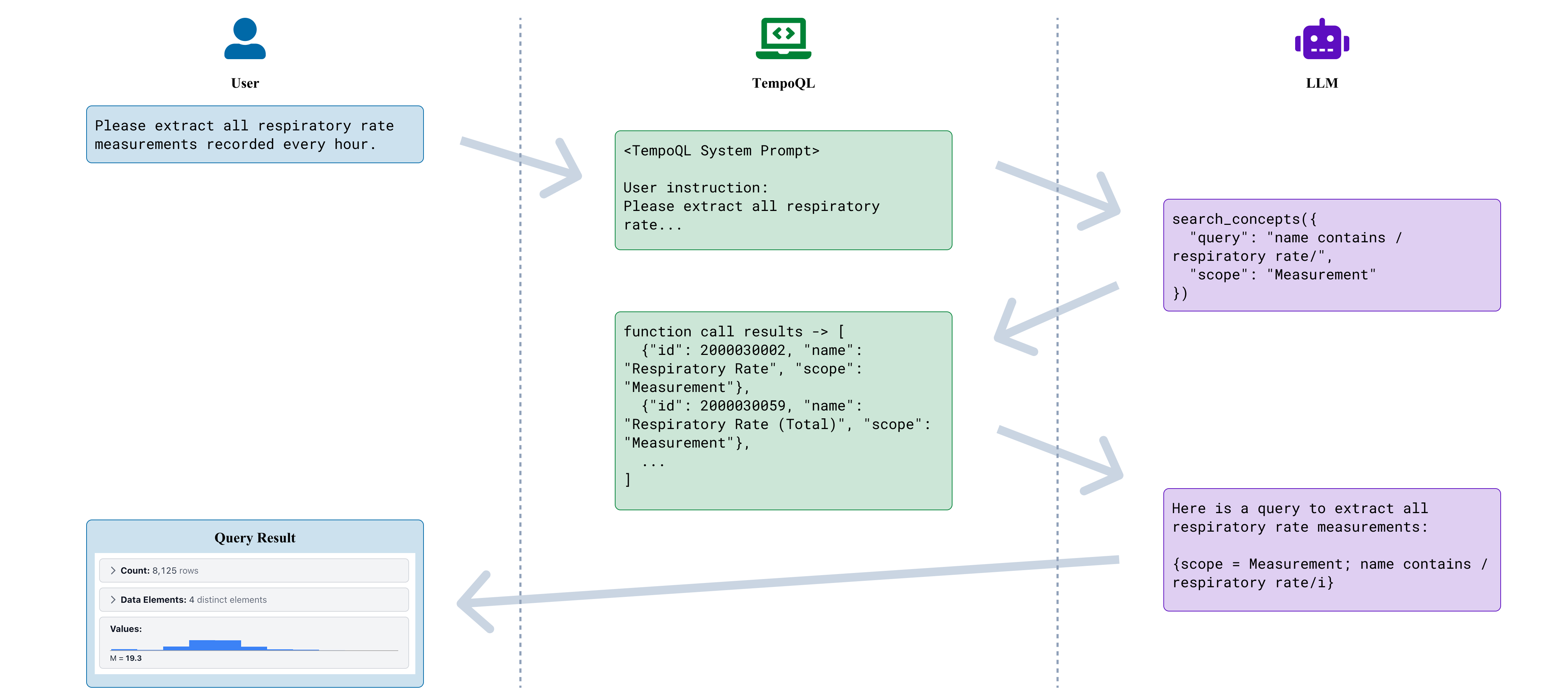}
    \caption{User - TempoQL - LLM interaction workflow showing how a user’s natural language request is translated into a structured TempoQL query through LLM-assisted concept search.}
    \label{fig:function-calling-flow}
\end{figure*}

Fig. \ref{fig:function-calling-flow} illustrates the process of translating a natural language query into an executable TempoQL query through LLM. The user request \texttt{"extract all respiratory rate measurements recorded every hour"} is passed through the TempoQL system prompt, which parsed into structured prompt to the LLM. The LLM identifies relevant measurement concepts (e.g., “Respiratory Rate”) via the \texttt{search\_concepts} call and generates the corresponding structured query \texttt{\{scope = Measurement; name contains /respiratory rate/i\}}. The query is then executed, and the extracted data is returned to the user.

\subsection{Dataset Specification}
\label{apd:dataset-spec}

As described in Sec. \ref{sec:data-elements}, TempoQL uses a Dataset Specification to map database tables and column names to the fields required for its core types.
The specification comprises three sections: (1) a list of available data element tables and how to map their columns to Attributes, Events, and/or Intervals; (2) a list of vocabulary tables that can map to concept IDs in the data element tables; and (3) optional joins to add fields to a data element table from another table.
Joins can be used to create consistent primary keys across tables.
For example, the abbreviated Dataset Specification below configures TempoQL to read from MIMIC-IV, where each ICU stay (defined by the \texttt{stay\_id} column) is an independent trajectory.
Tables that do not have a \texttt{stay\_id} column are specified to be joined on the \texttt{icustays} table before executing queries.

Dataset Specifications can be easily modified in Python code while setting up TempoQL for specific use cases, e.g. variations of OMOP.
They are designed to be sufficiently general that they could be written once for a given database and then reused for many analyses.
These specifications also serve a dual purpose as context for the LLM authoring tool; for instance, comments can be embedded to inform the model what additional fields are present and what formats the values take.

\subsubsection{Example: MIMIC-IV Dataset Specification}
\begin{small}
\begin{verbatim}
# MIMIC-IV is loaded from the physionet-data schema
PREFIX = "physionet-data.mimiciv_3_1"

# The available data element tables
TABLES = [
  {
    "source": PREFIX + "_hosp.admissions",
    "id_field": "stay_id",
    "scope": "Patient",
    "attributes": {
      "Marital Status": {
        "value_field": "marital_status"
      },
      "Race": {
        "value_field": "race"
      },
      "Hospital Mortality": {
        "value_field": "hospital_expire_flag"
      }
    }
  },
  {
    "source": PREFIX + "_hosp.patients",
    "id_field": "stay_id",
    "scope": "Patient",
    "attributes": {
      "Gender": {
        "value_field": "gender"
      },
      "Anchor Age": {
        "value_field": "anchor_age"
      },
      "Anchor Year": {
        "value_field": "anchor_year",
        "value_transform": "<function>"
      },
      "Date of Death": {
        "value_field": "dod"
      }
    },
    "comment": "All dates in the database have 
      been shifted to protect patient 
      confidentiality. Dates will be internally 
      consistent for the same patient, but 
      randomly distributed in the future. Dates 
      of birth which occur in the present time are 
      not true dates of birth. We can assume that 
      the patient's age at the attribute value 
      Anchor Year is the Anchor Age."
  },
  {
    "source": PREFIX + "_hosp.prescriptions",
    "type": "interval",
    "id_field": "stay_id",
    "interval_type_field": "drug",
    "start_time_field": "starttime",
    "end_time_field": "stoptime",
    "default_value_field": "dose_val_rx",
    "scope": "Medication",
    "comment": "The interval type is the name of 
      the drug. Value field 'dose_unit_rx' 
      represents the unit of the dose value; 
      'route' represents the way the drug is 
      administered."
  },
  {
    "source": PREFIX + "_icu.chartevents",
    "type": "event",
    "id_field": "stay_id",
    "time_field": "charttime",
    "concept_id_field": "itemid",
    "default_value_field": "value",
    "scope": "chartevents",
    "comment": "If a chart event sometimes has 
      string values returned, use value field 
      `valuenum' to specify that only numeric 
      results should be returned."
  },
  {
    "source": PREFIX + "_hosp.labevents",
    "type": "event",
    "id_field": "stay_id",
    "time_field": "charttime",
    "concept_id_field": "itemid",
    "default_value_field": "valuenum",
    "scope": "Lab",
    "comment": "If a lab test has string values, 
      use value field 'value' to return the 
      strings. By default only numeric values 
      are returned."
  },
  # ...
]

# Concept tables that map to one or more scopes
VOCABULARIES = [
  {
    "source": PREFIX + "_hosp.d_labitems",
    "concept_id_field": "itemid",
    "concept_name_field": "label",
    "scope": "Lab"
  },
  {
    "source": PREFIX + "_icu.d_items",
    "concept_id_field": "itemid",
    "concept_name_field": "label",
    "scope_field": "linksto",
    "scopes": [
      "chartevents",
      "inputevents",
      "outputevents",
      "procedureevents"
    ]
  }
]

# Joins to add fields (e.g. trajectory IDs) to
# data element tables before querying on them
JOINS = {
  PREFIX + "_hosp.admissions": {
    "dest_table": PREFIX + "_icu.icustays",
    "join_key": "hadm_id"
  },
  PREFIX + "_hosp.labevents": {
    "dest_table": PREFIX + "_icu.icustays",
    "join_key": "hadm_id"
  },
  # ...
}
\end{verbatim}
\end{small}

\section{Performance Evaluation Details}
\label{apd:perf-eval}
The query descriptions and code used throughout the Performance Evaluation (Sec. \ref{sec:perf-eval}) are presented in Tables \ref{tab:performance-eval-queries-1}, \ref{tab:performance-eval-queries-2}, \ref{tab:performance-eval-queries-3}, and \ref{tab:performance-eval-queries-4}.
This code is also provided in the GitHub repository.

For the execution time evaluation, we ran each TempoQL and BigQuery SQL query on a single machine (2023 MacBook Pro, 10-core M2 Pro CPU) over subsets of 1K, 5K, 10K, and 50K ICU stays in the MIMIC-IV dataset (result tables range in length from 300 to 4M rows).
We ran three identical replicates for each condition. 
The only difference between the TempoQL and SQL conditions in this experiment were that TempoQL retrieves results for individual data elements (e.g. Heart Rate) and performs aggregations locally, while the SQL condition executes the complete query in the database and returns the final results.

For the LLM query generation quality evaluation, we used the ``Prompt Instruction'' texts listed in the tables below as user instructions to Gemini 2.5 Pro.
We prefixed each prompt with a system prompt tailored to either TempoQL (see Appendix \ref{apd:llm-prompts}) or BigQuery SQL, then queried Gemini 10 times to assess how often it generated queries that could successfully extract data matching one of our reference solutions.
The reference solutions included the results returned by both our handwritten TempoQL and SQL queries, as well as up to four additional variants to allow for slightly different interpretations of the instruction.

\begin{table*}[]
    \centering
    \small
    \begin{tabular}{p{2.2cm}p{4cm}p{4cm}p{6cm}}
\toprule
\textbf{Query Name} & \textbf{Prompt Instruction} & \textbf{TempoQL} & \textbf{BigQuery SQL} \\ \midrule
Attributes & 
Extract each patient's age at the time of their ICU admission. & 
\begin{minipage}[t]{4cm}
\begin{scriptsize}
\begin{verbatim}({Admit Time} - {Anchor Year})
  as years
  + {Anchor Age}
\end{verbatim}\end{scriptsize}
\end{minipage} & 
\begin{minipage}[t]{6cm}
\begin{tiny}
\begin{verbatim}SELECT 
  stays.stay_id AS stay_id, 
  EXTRACT(YEAR FROM (stays.intime - 
    DATETIME(CONCAT(CAST(pat.anchor_year AS STRING), 
                    "-01-01"))))
    + pat.anchor_age AS age
FROM stays
INNER JOIN 
  `physionet-data.mimiciv_3_1_hosp.patients` pat
ON pat.subject_id = stays.subject_id
ORDER BY stay_id ASC

\end{verbatim}\end{tiny}
\end{minipage} \\ \midrule

Events &
Extract all respiratory rate measurements from the chartevents table. & 
\begin{minipage}[t]{4cm}
\begin{scriptsize}
\begin{verbatim}{Respiratory Rate; 
 scope = chartevents}
\end{verbatim}\end{scriptsize}
\end{minipage} & 
\begin{minipage}[t]{6cm}
\begin{tiny}
\begin{verbatim}WITH matching_eventids AS (
  SELECT DISTINCT 
    d.itemid AS itemid 
  FROM `physionet-data.mimiciv_3_1_icu.d_items` d
  WHERE d.label = 'Respiratory Rate'
)
SELECT 
  ce.stay_id AS stay_id, 
  ce.charttime AS time, 
  ce.itemid AS eventtype,
  ce.value AS value
FROM `physionet-data.mimiciv_3_1_icu.chartevents` ce
INNER JOIN stays
ON ce.stay_id = stays.stay_id
INNER JOIN matching_eventids 
ON ce.itemid = matching_eventids.itemid
ORDER BY stay_id, time ASC

\end{verbatim}\end{tiny}
\end{minipage} \\ \midrule

String Operations &
Extract a boolean value for each diagnosis indicating whether it is related to diabetes. ICD-9/10 codes related to diabetes start with the following possible prefixes: 401, 402, 403, 404, 405, I10, I11, I12, I13, I15. Use the ICU discharge time as the timestamp for the diagnosis if applicable. & 
\begin{minipage}[t]{4cm}
\begin{scriptsize}
\begin{verbatim}
{Diagnosis; scope = Diagnosis} 
  contains 
  /\\b(?:40[1-5]|I1[01235])/i
\end{verbatim}\end{scriptsize}
\end{minipage} & 
\begin{minipage}[t]{6cm}
\begin{tiny}
\begin{verbatim}
SELECT
  vas.stay_id,
  vas.outtime AS time,
  'Diagnosis' AS type,
  CAST(REGEXP_CONTAINS(
    dia.icd_code, 
    r"(?i)\b(?:40[1  5]|I1[01235])") 
    AS INT64) AS value
FROM
  `physionet-data.mimiciv_3_1_hosp.diagnoses_icd` AS dia
INNER JOIN
  stays AS vas
  ON dia.hadm_id = vas.hadm_id
ORDER BY
  stay_id ASC,
  time ASC

\end{verbatim}\end{tiny}
\end{minipage} \\ \midrule

Discretizing Observations &
Extract all Platelet Count observations from the lab results table, without excluding those with missing values. While preserving missingness, discretize the values so that if value $< 130$, the output value is 'Low', and if value $\geq 400$ the output is 'High', otherwise it should be 'Normal'. & 
\begin{minipage}[t]{4cm}
\begin{scriptsize}
\begin{verbatim}
{Platelet Count; 
 scope = Lab; 
 value = valuenum} 
cut bins 
  [-inf, 130, 400, inf] 
named 
  ['Low', 'Normal', 'High']
\end{verbatim}\end{scriptsize}
\end{minipage} & 
\begin{minipage}[t]{6cm}
\begin{tiny}
\begin{verbatim}
WITH matching_eventids AS (
  SELECT DISTINCT d.itemid AS itemid 
  FROM `physionet-data.mimiciv_3_1_hosp.d_labitems` d
  WHERE d.label = 'Platelet Count'
)
SELECT
  s.stay_id,
  le.charttime AS time,
  'Platelet Count' AS eventtype,
  CASE
    WHEN le.valuenum < 130 THEN 'Low'
    WHEN le.valuenum BETWEEN 130 AND 400 THEN 'Normal'
    ELSE 'High'
  END AS value
FROM
  `physionet-data.mimiciv_3_1_hosp.labevents` AS le
INNER JOIN
  `stays` AS s
  ON le.hadm_id = s.hadm_id 
  AND le.subject_id = s.subject_id
INNER JOIN
  `matching_eventids` AS mei
  ON le.itemid = mei.itemid
ORDER BY
  s.stay_id,
  le.charttime

\end{verbatim}\end{tiny}
\end{minipage}\\

\bottomrule
    \end{tabular}
    \caption{Performance evaluation queries, part 1.}
    \label{tab:performance-eval-queries-1}
\end{table*}

\begin{table*}[]
    \centering
    \small
    \begin{tabular}{p{2.2cm}p{4cm}p{4cm}p{6cm}}
\toprule
\textbf{Query Name} & \textbf{Prompt Instruction} & \textbf{TempoQL} & \textbf{BigQuery SQL} \\ \midrule

Patient-Level Aggregation &
Provide the minimum value for the 'Non Invasive Blood Pressure mean' event from chartevents over each patient's entire record. & 
\begin{minipage}[t]{4cm}
\begin{scriptsize}
\begin{verbatim}
min 
  {Non Invasive Blood Pressure 
   mean; scope = chartevents} 
from #mintime to #maxtime
\end{verbatim}\end{scriptsize}
\end{minipage} & 
\begin{minipage}[t]{6cm}
\begin{tiny}
\begin{verbatim}
WITH matching_eventids AS (
  SELECT DISTINCT d.itemid AS itemid 
  FROM `physionet-data.mimiciv_3_1_icu.d_items` d
  WHERE d.label = 'Non Invasive Blood Pressure mean'
),
matching_events AS (
  SELECT DISTINCT 
    ce.stay_id AS stay_id, ce.charttime AS charttime,
    ce.value AS value
  FROM `physionet-data.mimiciv_3_1_icu.chartevents` ce
  INNER JOIN matching_eventids 
  ON ce.itemid = matching_eventids.itemid
)
SELECT DISTINCT 
  stays.stay_id AS stay_id, 
  MIN(matching_events.value) AS value
FROM stays 
LEFT JOIN matching_events
ON matching_events.stay_id = stays.stay_id
GROUP BY stays.stay_id ORDER BY stay_id ASC

\end{verbatim}\end{tiny}
\end{minipage}\\ \midrule

Daily Aggregation &
Write a query that returns a row for every day in the patient's admission starting from the ICU admission time to the discharge time. Each row's value should contain the average lactate value in the preceding 24 hours. & 
\begin{minipage}[t]{4cm}
\begin{scriptsize}
\begin{verbatim}
mean 
  {Lactate; scope = Lab; 
   value = valuenum} 
from #now - 1 day to #now 
every 1 day 
  from {Admit Time} 
  to {Discharge Time}
\end{verbatim}\end{scriptsize}
\end{minipage} & 
\begin{minipage}[t]{6cm}
\begin{tiny}
\begin{verbatim}
WITH matching_eventids AS (
  SELECT DISTINCT d.itemid AS itemid 
  FROM `physionet-data.mimiciv_3_1_hosp.d_labitems` d
  WHERE d.label = 'Lactate'
),
DailyTimePoints AS (
  SELECT swh.hadm_id, swh.stay_id,
    generated_time AS time_point_end_window
  FROM
    stays AS swh,
    UNNEST(GENERATE_TIMESTAMP_ARRAY(
      CAST(swh.intime AS TIMESTAMP),
      CAST(swh.outtime AS TIMESTAMP),
      INTERVAL 24 HOUR
    )) AS generated_time
),
LactateMeasurements AS (
  SELECT le.hadm_id, le.charttime, le.valuenum
  FROM `physionet-data.mimiciv_3_1_hosp.labevents` AS le
  INNER JOIN matching_eventids AS li
    ON le.itemid = li.itemid
)
SELECT DISTINCT
  dtp.stay_id, dtp.time_point_end_window AS time,
  AVG(lm.valuenum) AS value
FROM
  DailyTimePoints AS dtp
LEFT JOIN
  LactateMeasurements AS lm
  ON dtp.hadm_id = lm.hadm_id
  AND CAST(lm.charttime AS TIMESTAMP) >= 
    TIMESTAMP_SUB(dtp.time_point_end_window, INTERVAL 24 HOUR)
  AND CAST(lm.charttime AS TIMESTAMP) < 
    dtp.time_point_end_window
GROUP BY dtp.stay_id, dtp.time_point_end_window
ORDER BY dtp.stay_id, dtp.time_point_end_window

\end{verbatim}\end{tiny}
\end{minipage}\\ \midrule

Aggregation in Overlapping Intervals &
Write a query that returns a row for every 4 hours in the patient's admission starting from the ICU admission time to the ICU discharge time. Each row's value should contain the minimum value of the mean blood pressure in the preceding 8 hours. & 
\begin{minipage}[t]{4cm}
\begin{scriptsize}
\begin{verbatim}
min 
  {Non Invasive Blood Pressure 
   mean; scope = chartevents} 
from #now - 8 h to #now 
every 4 h 
  from {Admit Time} 
  to {Discharge Time}
\end{verbatim}\end{scriptsize}
\end{minipage} & 
\begin{minipage}[t]{6cm}
\begin{tiny}
\begin{verbatim}
WITH matching_eventids AS (
  SELECT DISTINCT d.itemid AS itemid 
  FROM `physionet-data.mimiciv_3_1_icu.d_items` d
  WHERE d.label = 'Non Invasive Blood Pressure mean'
), GeneratedTimePoints AS (
  SELECT s.stay_id, generated_time AS time_point_end_window
  FROM
    `stays` AS s,
    UNNEST(GENERATE_TIMESTAMP_ARRAY(
      CAST(s.intime AS TIMESTAMP),
      CAST(s.outtime AS TIMESTAMP) ,
      INTERVAL 4 HOUR
    )) AS generated_time
),
MBP_Measurements AS (
  SELECT ce.stay_id, ce.charttime, ce.valuenum
  FROM `physionet-data.mimiciv_3_1_icu.chartevents` AS ce
  INNER JOIN `matching_eventids` AS mei
    ON ce.itemid = mei.itemid
)
SELECT DISTINCT
  gtp.stay_id, gtp.time_point_end_window AS time,
  MIN(mbp.valuenum) AS value
FROM
  GeneratedTimePoints AS gtp
LEFT JOIN
  MBP_Measurements AS mbp
  ON gtp.stay_id = mbp.stay_id
  AND CAST(mbp.charttime AS TIMESTAMP) >= 
    TIMESTAMP_SUB(gtp.time_point_end_window, INTERVAL 8 HOUR)
  AND CAST(mbp.charttime AS TIMESTAMP) < 
    gtp.time_point_end_window
GROUP BY gtp.stay_id, gtp.time_point_end_window
ORDER BY gtp.stay_id, gtp.time_point_end_window

\end{verbatim}\end{tiny}
\end{minipage}\\

\bottomrule
    \end{tabular}
    \caption{Performance evaluation queries, part 2.}
    \label{tab:performance-eval-queries-2}
\end{table*}

\begin{table*}[]
    \centering
    \small
    \begin{tabular}{p{2.2cm}p{4cm}p{4cm}p{6cm}}
\toprule
\textbf{Query Name} & \textbf{Prompt Instruction} & \textbf{TempoQL} & \textbf{BigQuery SQL} \\ \midrule

Aggregating Existence at Event Times &
Write a query that returns a row at every start of an invasive ventilation event from the procedures table. Use the specific event called 'Invasive Ventilation'. Each row's value should contain a boolean value indicating if there was a previous invasive ventilation event for this ICU stay. & 
\begin{minipage}[t]{4cm}
\begin{scriptsize}
\begin{verbatim}
exists 
  {Invasive Ventilation; 
   scope = procedureevents} 
before #now 
at every start(
  {Invasive Ventilation; 
   scope = procedureevents}
)
\end{verbatim}\end{scriptsize}
\end{minipage} & 
\begin{minipage}[t]{6cm}
\begin{tiny}
\begin{verbatim}
WITH matching_eventids AS (
  SELECT DISTINCT d.itemid AS itemid 
  FROM `physionet-data.mimiciv_3_1_icu.d_items` d
  WHERE d.label = 'Invasive Ventilation'
),
VentilationEvents AS (
  SELECT ce.stay_id, ce.starttime
  FROM `physionet-data.mimiciv_3_1_icu.procedureevents` AS ce
  INNER JOIN `stays`
  ON ce.stay_id = stays.stay_id
  INNER JOIN `matching_eventids` AS mei
  ON ce.itemid = mei.itemid
)
SELECT
  ve.stay_id,
  ve.starttime AS time,
  CASE
    WHEN LAG(ve.starttime) OVER (
      PARTITION BY ve.stay_id ORDER BY ve.starttime
    ) IS NOT NULL THEN 1
    ELSE 0
  END AS value
FROM VentilationEvents AS ve
ORDER BY ve.stay_id, ve.starttime

\end{verbatim}\end{tiny}
\end{minipage}\\ \midrule

Aggregating Counts at Event Times &
Write a query that returns a row for every occurrence of a Heart Rhythm chart event. Each row's value should contain the count of all Cardioversion/Defibrillation procedure events that start within the 24 hours after the heart rhythm observation. & 
\begin{minipage}[t]{4cm}
\begin{scriptsize}
\begin{verbatim}
count 
  {Cardioversion/Defibrillation; 
   scope = procedureevents} 
from #now to #now + 24 h 
at every 
  {Heart Rhythm; 
   scope = chartevents}
\end{verbatim}\end{scriptsize}
\end{minipage} & 
\begin{minipage}[t]{6cm}
\begin{tiny}
\begin{verbatim}
WITH HeartRhythmItemIDs AS (
  SELECT itemid 
  FROM `physionet-data.mimiciv_3_1_icu.d_items`
  WHERE label = 'Heart Rhythm'
),
CardioDefibItemIDs AS (
  SELECT itemid 
  FROM `physionet-data.mimiciv_3_1_icu.d_items`
  WHERE label = 'Cardioversion/Defibrillation'
),
HeartRhythmEvents AS (
  SELECT ce.stay_id, ce.charttime, ce.value
  FROM `physionet-data.mimiciv_3_1_icu.chartevents` AS ce
  INNER JOIN stays AS s
  ON ce.stay_id = s.stay_id
  INNER JOIN HeartRhythmItemIDs AS hri
  ON ce.itemid = hri.itemid
),
CardioDefibProcedures AS (
  SELECT ce.stay_id, ce.starttime AS procedure_charttime
  FROM
    `physionet-data.mimiciv_3_1_icu.procedureevents` AS ce
  INNER JOIN stays AS s
  ON ce.stay_id = s.stay_id
  INNER JOIN CardioDefibItemIDs AS cdi
  ON ce.itemid = cdi.itemid
)
SELECT hre.stay_id, hre.charttime AS time,
  COUNT(cdp.procedure_charttime) AS value
FROM HeartRhythmEvents AS hre
LEFT JOIN
  CardioDefibProcedures AS cdp
  ON hre.stay_id = cdp.stay_id
  AND 
    CAST(cdp.procedure_charttime AS TIMESTAMP) 
    BETWEEN CAST(hre.charttime AS TIMESTAMP)
    AND TIMESTAMP_ADD(CAST(hre.charttime AS TIMESTAMP), 
      INTERVAL 24 HOUR)
GROUP BY hre.stay_id, hre.charttime, hre.value
ORDER BY hre.stay_id, hre.charttime

\end{verbatim}\end{tiny}
\end{minipage}\\ \midrule

Rolling Difference &
Write a query that returns a row for every occurrence of a Temperature Fahrenheit chart event. Each row's value should contain the difference between this temperature and the average of the temperature chart events for this patient in the last 8 hours. & 
\begin{minipage}[t]{4cm}
\begin{scriptsize}
\begin{verbatim}
temp - (
  mean
    temp 
  from #now - 8 h to #now 
  at every temp
) 
with temp as 
  {Temperature Fahrenheit; 
   scope = chartevents}
\end{verbatim}\end{scriptsize}
\end{minipage} & 
\begin{minipage}[t]{6cm}
\begin{tiny}
\begin{verbatim}
WITH matching_eventids AS (
  SELECT DISTINCT d.itemid AS itemid 
  FROM `physionet-data.mimiciv_3_1_icu.d_items` d
  WHERE d.label = 'Temperature Fahrenheit'
),
TemperatureEvents AS (
  SELECT ce.stay_id, ce.charttime, ce.valuenum
  FROM `physionet-data.mimiciv_3_1_icu.chartevents` AS ce
  INNER JOIN `stays` AS s
  ON ce.stay_id = s.stay_id
  INNER JOIN `matching_eventids` AS mei
  ON ce.itemid = mei.itemid
)
SELECT
  te.stay_id, te.charttime AS time,
  'Temperature' AS eventtype,
  te.valuenum - AVG(te.valuenum) OVER (
    PARTITION BY te.stay_id
    ORDER BY UNIX_SECONDS(CAST(te.charttime AS TIMESTAMP))
    RANGE BETWEEN 28800 PRECEDING AND 1 PRECEDING
  ) AS value
FROM TemperatureEvents AS te
ORDER BY te.stay_id, te.charttime

\end{verbatim}\end{tiny}
\end{minipage}\\

\bottomrule
    \end{tabular}
    \caption{Performance evaluation queries, part 3.}
    \label{tab:performance-eval-queries-3}
\end{table*}

\begin{table*}[]
    \centering
    \small
    \begin{tabular}{p{2.2cm}p{4cm}p{4cm}p{6cm}}
\toprule
\textbf{Query Name} & \textbf{Prompt Instruction} & \textbf{TempoQL} & \textbf{BigQuery SQL} \\ \midrule

Imputing Missing Values &
Write a query that returns a row every 4 hours starting from the ICU admission time to the ICU discharge time. Each row's value should contain the average Temperature Fahrenheit chart value in the preceding 4 hours, and if the value is missing then it should be the mean temperature value over all patients. & 
\begin{minipage}[t]{4cm}
\begin{scriptsize}
\begin{verbatim}
mean 
  {Temperature Fahrenheit; 
   scope = chartevents} 
from #now - 4 h to #now 
impute mean 
every 4 h 
  from {Admit Time} 
  to {Discharge Time}
\end{verbatim}\end{scriptsize}
\end{minipage} & 
\begin{minipage}[t]{6cm}
\begin{tiny}
\begin{verbatim}
WITH matching_eventids AS (
  SELECT DISTINCT d.itemid AS itemid 
  FROM `physionet-data.mimiciv_3_1_icu.d_items` d
  WHERE d.label = 'Temperature Fahrenheit'
),
OverallMeanTemperature AS (
  SELECT AVG(ce.valuenum) AS global_avg_temp
  FROM `physionet-data.mimiciv_3_1_icu.chartevents` AS ce
  INNER JOIN `matching_eventids` AS mei
  ON ce.itemid = mei.itemid
),
GeneratedTimePoints AS (
  SELECT s.stay_id, 
    generated_time AS time_point_end_window
  FROM `stays` AS s,
  UNNEST(GENERATE_TIMESTAMP_ARRAY(
    CAST(s.intime AS TIMESTAMP),
    CAST(s.outtime AS TIMESTAMP),
    INTERVAL 4 HOUR
  )) AS generated_time
),
FilteredTemperatureEvents AS (
  SELECT ce.stay_id, ce.charttime, ce.valuenum
  FROM `physionet-data.mimiciv_3_1_icu.chartevents` AS ce
  INNER JOIN `stays` AS s
  ON ce.stay_id = s.stay_id
  INNER JOIN `matching_eventids` AS mei
  ON ce.itemid = mei.itemid
  WHERE ce.valuenum IS NOT NULL
)
SELECT
  gtp.stay_id,
  gtp.time_point_end_window AS time,
  COALESCE(
    AVG(fte.valuenum),
    (SELECT global_avg_temp FROM OverallMeanTemperature)
  ) AS value
FROM GeneratedTimePoints AS gtp
LEFT JOIN FilteredTemperatureEvents AS fte
ON gtp.stay_id = fte.stay_id
AND CAST(fte.charttime AS TIMESTAMP) > 
  TIMESTAMP_SUB(gtp.time_point_end_window, INTERVAL 4 HOUR)
AND CAST(fte.charttime AS TIMESTAMP) <= 
  gtp.time_point_end_window
GROUP BY gtp.stay_id, gtp.time_point_end_window
ORDER BY gtp.stay_id, gtp.time_point_end_window

\end{verbatim}\end{tiny}
\end{minipage}\\ \midrule

Carrying Values Forward &
Write a query that returns a row for every 24 hours starting from the ICU admission time to the ICU discharge time. Each row's value should contain the EARLIEST observed value for the O2 delivery device (the chart event is called 'O2 Delivery Device(s)') in the preceding 24 hours. Values should be carried forward by up to 2 days if subsequent values are missing. & 
\begin{minipage}[t]{4cm}
\begin{scriptsize}
\begin{verbatim}
first 
  {O2 Delivery Device(s); 
   scope = chartevents} 
from #now - 1 day to #now 
carry 2 days 
every 1 day 
  from {Admit Time} 
  to {Discharge Time}
\end{verbatim}\end{scriptsize}
\end{minipage} & 
\begin{minipage}[t]{6cm}
\begin{tiny}
\begin{verbatim}
WITH matching_eventids AS (
  SELECT DISTINCT d.itemid AS itemid 
  FROM `physionet-data.mimiciv_3_1_icu.d_items` d
  WHERE d.label = 'O2 Delivery Device(s)'
),
GeneratedTimePoints AS (
  SELECT s.stay_id, 
    generated_time AS time_point_end_window
  FROM `stays` AS s,
    UNNEST(GENERATE_TIMESTAMP_ARRAY(
      CAST(s.intime AS TIMESTAMP),
      CAST(s.outtime AS TIMESTAMP),
      INTERVAL 24 HOUR                          
    )) AS generated_time
),
O2Events AS (
  SELECT ce.stay_id, ce.charttime, ce.value
  FROM `physionet-data.mimiciv_3_1_icu.chartevents` AS ce
  INNER JOIN `stays` AS s
  ON ce.stay_id = s.stay_id
  INNER JOIN `matching_eventids` AS mei
  ON ce.itemid = mei.itemid
  WHERE ce.value IS NOT NULL
),
WindowsWithEarliestValue AS (
  SELECT gtp.stay_id,
    gtp.time_point_end_window,
    ARRAY_AGG(
      o2e.value ORDER BY o2e.charttime ASC LIMIT 1
    )[OFFSET(0)] AS current_window_value
  FROM GeneratedTimePoints AS gtp
  LEFT JOIN O2Events AS o2e
  ON gtp.stay_id = o2e.stay_id
    AND CAST(o2e.charttime AS TIMESTAMP) > 
      TIMESTAMP_SUB(gtp.time_point_end_window, INTERVAL 24 HOUR)
    AND CAST(o2e.charttime AS TIMESTAMP) <= 
      gtp.time_point_end_window
  GROUP BY gtp.stay_id, gtp.time_point_end_window
)
SELECT wwev.stay_id,
  wwev.time_point_end_window AS time,
  LAST_VALUE(wwev.current_window_value IGNORE NULLS) OVER (
    PARTITION BY wwev.stay_id
    ORDER BY UNIX_SECONDS(wwev.time_point_end_window)
    RANGE BETWEEN 172800 PRECEDING AND 1 PRECEDING
  ) AS value
FROM
WindowsWithEarliestValue AS wwev
ORDER BY wwev.stay_id, wwev.time_point_end_window

\end{verbatim}\end{tiny}
\end{minipage}\\
\bottomrule

  \end{tabular}
  \caption{Performance evaluation queries, part 4.}
    \label{tab:performance-eval-queries-4}
\end{table*}

\section{Example Sepsis Queries}\label{apd:second}

We developed five queries to define sepsis patient cohorts, 108 queries for data extraction, and 224 queries for modeling.
The cohort definition and data extraction queries were adapted slightly for each database, while the modeling queries remained constant across the three databases.
In Table \ref{tab:sepsis-example-queries} we show examples of the queries from each category across the datasets.

\begin{table*}
\centering
\small
\begin{tabular}{p{5cm}p{5cm}p{5cm}}
\toprule
\multicolumn{3}{l}{\textbf{Cohort Selection}} \\
\midrule
\multicolumn{3}{l}{\textbf{Example:} Sepsis Diagnosis}\\
\textbf{MIMIC-IV} & \textbf{eICU} & \textbf{EHRSHOT}\\
\begin{minipage}[t]{\linewidth}
\footnotesize \begin{verbatim}
any (
  {Diagnosis; scope = Diagnosis} 
  contains /78552|99591|99592/
) 
from {Admit Time} 
to ({Discharge Time} 
    impute {Admit Time}) + 1 day
\end{verbatim}\vspace{2pt}
\end{minipage}
&
\begin{minipage}[t]{\linewidth}
\footnotesize \begin{verbatim}
any (
  {Diagnosis; scope = Diagnosis} 
  contains 
  /785\.52|995\.91|995\.92/
) 
from #mintime to #maxtime
\end{verbatim}\vspace{2pt}
\end{minipage}
&
\begin{minipage}[t]{\linewidth}
\footnotesize \begin{verbatim}
exists (
  {name startswith /sepsis/i; 
   scope = SNOMED}
) 
from time(Culture) - 1 day 
to time(Culture) + 30 days
\end{verbatim}\vspace{2pt}
\end{minipage}
\\ \midrule
\multicolumn{3}{l}{\textbf{Data Extraction}}\\ \midrule
\multicolumn{3}{l}{\textbf{Example:} Lactic Acid measurements}\\[2pt]
\textbf{MIMIC-IV} & \textbf{eICU} & \textbf{EHRSHOT}\\
\begin{minipage}[t]{\linewidth}
\footnotesize \begin{verbatim}
union(
  {Lactic Acid; 
   scope = chartevents}, 
  {Lactate; 
   scope = Lab; 
   value = valuenum}
)
\end{verbatim}\vspace{2pt}
\end{minipage}
&
\begin{minipage}[t]{\linewidth}
\footnotesize \begin{verbatim}
{lactate; scope = Lab}
\end{verbatim}\vspace{2pt}
\end{minipage}
&
\begin{minipage}[t]{\linewidth}
\footnotesize \begin{verbatim}
{name startswith 
 'Lactate [Moles/volume]'; 
 scope = LOINC; 
 value = numeric_value}
\end{verbatim}\vspace{2pt}
\end{minipage}
\\
\multicolumn{3}{l}{\textbf{Example:} Norepinephrine Administration}\\[2pt]
\textbf{MIMIC-IV} & \textbf{eICU} & \textbf{EHRSHOT}\\
\begin{minipage}[t]{\linewidth}
\footnotesize \begin{verbatim}
{Norepinephrine; 
 scope = inputevents; 
 value = amount} 
where {Norepinephrine; 
       scope = inputevents; 
       value = rateuom} = 
      'mcg/kg/min'
\end{verbatim}\vspace{2pt}
\end{minipage}
&
\begin{minipage}[t]{\linewidth}
\footnotesize Because start/end times of norepinephrine are not recorded in eICU, assume a 30-minute administration window to estimate totals.\par
\scriptsize
\begin{verbatim}
vals * 30 
with vals as union(
  ({name = 'Norepinephrine (mcg/min)'; 
    scope = Infusion} 
   where #value contains /^[-0-9.]+$/) 
   / Weight,
  ({name = 'Norepinephrine (mcg/kg/min)'; 
    scope = Infusion} 
   where #value contains /^[-0-9.]+$/),
  ({name = 'Norepinephrine (ml/hr)'; 
    scope = Infusion} 
   where #value contains /^[-0-9.]+$/) 
   / 60 / Weight
)
\end{verbatim}\vspace{2pt}
\end{minipage}
&
\begin{minipage}[t]{\linewidth}
\footnotesize Estimate amounts from RxNorm codes (infusion rate and endpoints not recorded).\par
\begin{verbatim}
union(
  assign({id = 'RxNorm/242969'; 
          scope = RxNorm}, 
         4 * 1 / Weight),
  assign({id = 'RxNorm/2475337'; 
          scope = RxNorm}, 
         250 * 0.01 / Weight)
) * 1000
\end{verbatim}\vspace{2pt}
\end{minipage}
\\[2pt] \midrule
\multicolumn{3}{l}{\textbf{Model Input Features}}\\ \midrule
\multicolumn{3}{l}{\textbf{Example:} Lactic Acid}\\
\multicolumn{3}{l}{%
\footnotesize
\texttt{last LacticAcid from \#now - 4 h to \#now carry 24 hours}
}\\[2pt]
\multicolumn{3}{l}{\textbf{Example:} Norepinephrine}\\
\multicolumn{3}{l}{%
\footnotesize
\texttt{sum Norepinephrine from \#now - 4 h to \#now impute 0}
}\\
\bottomrule
\end{tabular}
\caption{Example queries used in the sepsis prediction model case study.}
    \label{tab:sepsis-example-queries}
\end{table*}

\section{LLM Prompts and TempoQL Reference}\label{apd:llm-prompts}

This appendix provides the prompts used to guide large language models in generating, explaining, and debugging TempoQL queries in our system.
These prompts also double as a reference for the available query language syntax features.

Since TempoQL would not have been present in our LLM's training corpus, the following base prompt was used to provide it with the structure of TempoQL syntax via an in-context learning strategy:

\begin{small}
\begin{verbatim}
You are a helpful data analysis assistant.
You can help users understand their data
and write queries. You are an expert on a
new query language called TempoQL that is
specialized to deal with electronic health
record data. Below is some information
about how TempoQL works:

---

Datasets in TempoQL contain one or more
**scopes**. Each scope contains one or
more data elements, which can take one of
three types: attributes (do not change
throughout a patient trajectory), events
(associated with a single point in time),
or intervals (occur between two points in
time). The available scopes are defined
in the Table Context below. Some scopes
have pre-defined attributes, events, or
interval types, while others refer to
concepts using a clinical vocabulary. You
can query for concepts that match a
particular data element query using the
`search_concepts' function.

In Tempo's syntax, you begin by selecting
data elements from this underlying dataset
using **data element queries**. Data
element queries are wrapped in curly
braces and consist of one or more
components separated by semicolons, where
each component operates on a `field'
(`id' - concept ID, `name' - concept name,
`type' - attributes/events/intervals,
`scope' - scope in the dataset, or
`value' - field to use for the result),
and looks like this:

- `<field> = <value>' (constrain to only
one value)
- `<field> in (<value 1>, <value 2>, ...)'
(constrain to one of a set of values, not
valid for type, scope, or value)
- `<field> contains <pattern>' , or
`matches', `startswith', `endswith'
(constrain to values that match a given
regex pattern, not valid for type, scope,
or value)

For example, to get observations for
respiratory rate, the following queries
could work:

- `{name in ("Breath Rate", "Respiratory
Rate", "Resp Rate", "Resp Rate (Total)")}'
- `{scope = observations; id in (220210,
3337, 224422, 618, 3603, 615)}'
- `{scope = observations; name contains
/(breath|resp\w+) rate/I}'

You can also use a shorthand for
extracting a single concept by name by
removing the `name = ' portion like so:
`{Respiratory Rate; scope = observations}'.

Data element queries can then be operated
on using arithmetic and logical operations
similar to SQL. By default, operations
apply to the value associated with a data
field (such as a temperature measurement).
The following examples demonstrate valid
TempoQL syntax:

- `({Temperature Fahrenheit} - 32) *
5 / 9'
- `case when {Gender} = 'Female' then 2
else 1 end'
- `{Weight} / ({Height} ^ 2)'

An important part of TempoQL syntax is
**aggregations**, which result in a Time
Series that is aligned to a user-defined
*timestep definition*. The timestep
definition specifies the bounds and
frequency of timesteps that should be
selected within each trajectory. Timestep
definitions always take one of the
following formats:

- `at every <event>' (optionally provide
time bounds: `from <start time> to
<end time>')
- `every <duration>' (optionally provide
time bounds)
- `at [<list of times>]'

Durations are expressed as a number plus
an English unit of time, e.g. `4 hours',
`3 days', `30 seconds', `1 hour'.

Each aggregation is computed over the
observations that fall within the provided
*aggregation bounds*, which are typically
a function of the special marker `#now'
(which represents the time of the current
timestep). For example, the aggregation
bounds `from #now - 4 hours to #now' would
extract observations in the 4 hours
preceding each time in the timestep
definition. The shorthands `before <time>',
`after <time>', and `at <time>' are
available. The markers `#mintime' and
`#maxtime' can be used to represent the
earliest and latest observations for each
patient.

The aggregation function to be applied
over all observations must be one of the
following: sum, mean, median, min, max,
first, last, any, all, all nonnull,
exists, exists nonnull, count, count
distinct, count distinct nonnull, count
nonnull, or integral. Each function always
produces a single scalar value over all
matching observations.

If the value being aggregated is an
interval, you can specify if the value
should be treated as a `rate', `amount',
`duration', or `value' by including that
keyword after the aggregation function.
For example, given an interval representing
the rate of an IV infusion called `{IV
Infusion}', `integral rate {IV Infusion}'
would compute the total amount of IV fluid
within the aggregation bounds.

Below are some examples of aggregation
expressions and what they do:
- `last {Temperature Celsius} from #now - 1
hour to #now every 4 hours from {Admission
Time} to {Discharge Time}': assumes there
are attributes called Admission Time and
Discharge Time, and computes the most
recent temperature in the last hour at
4-hour timepoints.
- `mean {Heart Rate} from #now - 8 hours to
#now every 1 day' - calculates the average
heart rate over the last 8 hours at
timepoints 1 day apart from each patient's
earliest observation to their latest.
- `exists {name contains /heart disease/i;
scope = Diagnosis} before #now at every
{Admission}' - assumes there is an event
called Admission, and looks for diagnosis
events related to heart disease before
each admission event.

The following additional features are
available in TempoQL:
- String and regex operations (contains,
startswith, endswith), e.g. `{Heart Rhythm}
contains /fibrillation/i'
- Filtering: `{Temperature Celsius} where
#value < 50'
- Carrying forward values in case of
missingness: `mean {Glucose; scope = Lab}
from #now - 8 hours to #now carry 12 hours
every 4 hours'
- Imputing missing values: `mean {Hemoglobin;
scope = Lab} before #now impute mean every
3 days'

You can use the following functions to
transform values:
- time(<Events>): returns a new Events
series where the values are the times of
each row.
- type(<Events|Intervals>): returns a new
Events or Intervals where the values are
the event/interval types (data elements).
- start(<Intervals>), end(<Intervals>):
returns a new Events for either the start
time or the end time of the given Intervals
- starttime(<Intervals>),
endtime(<Intervals>),
duration(<Intervals>): returns an Events
object containing the start times, the end
times, or the durations of each interval
- intervals(<Events>, <Events>): create an
Intervals object with the start and end
times specified by the two Events objects
- abs, max, min: standard math functions
- extract(expr, pattern) or extract(expr,
pattern, index): regex extraction
- union(a, b, ...): combine two or more
sets of Events or Intervals together
- assign(expr, value): assign the given
value to the elements in the expression,
broadcasting if necessary

---

Here is the Table Context for the dataset
you will be working with:

```
<DATASET_INFO>
'''

You will not have access to the data
itself, only this specification and the
names and IDs of concepts/data elements
stored within the data. When appropriate,
acknowledge that you cannot access the
underlying data for privacy reasons.

\end{verbatim}
\end{small}

The \texttt{<DATASET\_INFO>} variable in the prompt will be replaced by the Dataset Specification, an example of which is shown in Appendix \ref{apd:dataset-spec}.

We then add instructions for query generation, where the LLM translates plain human input into TempoQL queries:
\begin{small}
\begin{verbatim}
 Given this information, I will provide you 
with an instruction on a query to write. 
You may call the search_concepts function 
to retrieve a list of matching concepts, 
if needed. Remember that the dataset may 
not contain any of the event types used 
in the examples above. I recommend 
calling the search_concepts function 
one or more times and searching broadly, 
such as by using a case-insensitive regular
expression, since concept names may not 
match your initial search. You may then 
need to refine your concept query to 
select only the relevant concepts from 
the ones that are returned. Think 
carefully to ensure that the final query 
is simple but returns the most relevant 
data elements.

After retrieving any needed concepts, 
write a TempoQL query obeying the syntax 
description above. Your output should 
contain one or more multiline code blocks 
with the language 'tempoql' that contains 
your answer, as well as short explanations 
of how the query works at a level that a 
non-programmer expert on clinical data 
could understand. Only provide multiple 
options if the instruction I give you 
is ambiguous as to what query might 
be needed.

Instruction: <INSTRUCTION>
\end{verbatim}
\end{small}

% The \texttt{<INSTRUCTION>} variable here means the user questions, such as 
% \begin{small}
% \begin{verbatim}
% "Can you generate the query to 
% extract the average patient hearts
% from the last 90 days?",

% "Find all patients who were diagnosed 
% with diabetes before 2020 and who 
% were prescribed insulin within 
% one year after their diagnosis.",
% ...
% \end{verbatim}
% \end{small}

For explanation tasks, here is the prompt we extended on the base prompt to ask the LLM to describe in plain language what a given TempoQL query is intended to do.

\begin{small}
\begin{verbatim}
Given this information, I have written 
a TempoQL query below and I would like 
you to explain what it does.
You may call the search_concepts function 
to explain the meaning of data element 
queries if appropriate (for instance, 
to decode data elements referred to 
by a concept ID).

Be clear, concise and friendly but 
professional in your response, 
and do not include praise.

Provide a list of intuitive steps that 
the query follows to produce the response.
Some steps might include:
1. Data that the query extracts from 
the dataset
2. Transformations to the data
3. Aggregations used to structure the data
Include only the steps that actually exist 
in the query.
\end{verbatim}
\end{small}

For debugging tasks, we add the following prompt to instruct the LLM to identify and correct syntax or logical errors in user-written TempoQL queries, and to provide explanations of the fixes.

\begin{small}
\begin{verbatim}
Given this information, I have written a 
TempoQL query below which produced an error 
when I ran it. 
The error will be provided below the query 
and I would like you to explain the error 
and attempt to fix the issue. If you can 
fix the issue, provide the code in a code 
block labeled tempoql, like so:

```tempoql
tempo code goes here
'''

Make sure that the new query:
- Fixes any syntax or logical errors
- Uses correct data element references
- Follows proper TempoQL structure
- Is likely to execute successfully

Be clear, concise and friendly but 
professional, and do not include praise.
\end{verbatim}
\end{small}

\end{document}